\documentclass[aps,twocolumn,preprintnumbers,amsmath,amssymb,showpacs,nofootinbib,longbibliography]{revtex4-1}

\usepackage[english]{babel}
\usepackage[utf8]{inputenc}
\usepackage{amssymb}
\usepackage{amsmath}
\usepackage{color}
\usepackage{hyperref}
\usepackage{bbm}
\usepackage{epsfig}

\newcommand{\ket}[1]{|#1\rangle}

\newcommand{\expect}[1]{\langle #1 \rangle}
\newcommand{\zub}[2]{\langle \! \langle #1 \vert #2 \rangle\!\rangle}

\newcommand{\e}{\varepsilon}
\newcommand{\G}{\Gamma}
\newcommand{\s}{\sigma}
\renewcommand{\d}{\delta}

\newcommand{\TA}{T_A(\omega)}
\newcommand{\TAD}{T_A^{\rm DAR}(\omega)}
\newcommand{\TAC}{T_A^{\rm CAR}(\omega)}

\newcommand{\A}{\mathcal{A}}

\newcommand{\TKt}{T_{\rm K}^{SU(2)}}
\newcommand{\TKf}{T_{\rm K}^{SU(4)}}

\newcommand{\up}{\uparrow}
\newcommand{\down}{\downarrow}

\newcommand{\beq}{ \begin{equation} }
\newcommand{\eeq}{ \end{equation} }
\newcommand{\beqa}{\begin{eqnarray}}
\newcommand{\eeqa}{\end{eqnarray}}

\newcommand{\fig}[1]{Fig.~\ref{#1}}

\newcommand{\eq}[1]{Eq.~(\ref{#1})}

\hypersetup{
    colorlinks=true,        
    linkcolor=blue,
    citecolor=blue,
    filecolor=blue,
    urlcolor=blue
}

\begin{document}

\title{Kondo physics in double quantum dot based Cooper pair splitters}

\author{Kacper Wrze\'sniewski}
\affiliation{Faculty of Physics, Adam Mickiewicz University,
			 ul. Umultowska 85, 61-614 Pozna{\'n}, Poland}

\author{Ireneusz Weymann}
\email{weymann@amu.edu.pl}
\affiliation{Faculty of Physics, Adam Mickiewicz University,
			 ul. Umultowska 85, 61-614 Pozna{\'n}, Poland}
\date{\today}

\begin{abstract}
The Andreev transport properties of double quantum dot based Cooper pair splitters
with one superconducting and two normal leads are studied theoretically
in the Kondo regime. The influence of the superconducting pairing correlations
on the local density of states, Andreev transmission coefficient and 
Cooper pair splitting efficiency is thoroughly analyzed. 
It is shown that finite superconducting pairing potential quickly suppresses
the $SU(2)$ Kondo effect, which can however reemerge
for relatively large values of coupling to superconductor.
In the $SU(4)$ Kondo regime, a crossover
from the $SU(4)$ to the $SU(2)$ Kondo state
is found as the coupling to superconductor is enhanced.
The analysis is performed by means of the density-matrix
numerical renormalization group method.
\end{abstract}



\maketitle


\section{Introduction}
\label{sec:intro}


Creation, manipulation and detection of entangled pairs of electrons
is an important requirement for engineering quantum 
information and computation protocols in solid state systems
\cite{Loss1998Jan,Ruggiero2006,Nielsen2011Jan}.
As a natural source of entangled electrons one can consider superconductors,
in which two electrons with opposite spins form spin singlet states -- the Cooper pairs
\cite{Recher2001Apr,Tinkham2004Jun,Ciudad2015May}.
It has been demonstrated experimentally that it is possible
to extract and split Cooper pairs in a double quantum dot (DQD) setup
involving one superconductor (SC) and two normal (N) leads,
each attached to different quantum dot
\cite{Hofstetter2009Oct,Herrmann_PhysRevLett.104.026801,
Hofstetter_PhysRevLett.107.136801,Schindele_PhysRevLett.109.157002,
Das2012Nov,Fulop_PhysRevB.90.235412,
Fulop_PhysRevLett.115.227003,Tan_PhysRevLett.114.096602}.
In such a Cooper pair splitter (CPS), 
when the bias voltage $eV$ applied between the SC and N leads
is smaller than the superconducting energy gap $\Delta$,
the current flows through the system due to the Andreev reflection processes
\cite{andreev}.
One can generally distinguish two types of such processes:
(i) direct Andreev reflection (DAR), in which the Cooper pair
electrons tunnel through one arm of the device
and (ii) crossed Andreev reflection (CAR), when 
the Cooper pair electrons become split and 
each electron leaves the superconductor through different arm of the device
\cite{Beckmann_PhysRevLett.93.197003,Gramisch_PhysRevLett.115.216801}.
Since the latter processes are crucial for the creation of entangled electrons,
it is important to optimize the splitting efficiency $\eta$
of the device, i.e. to enhance the rate of CAR processes 
as compared to the DAR processes.
This can be obtained, for example, by tuning the position of the DQD's energy levels
and setting the system in an appropriate transport regime
\cite{Hofstetter2009Oct,Schindele_PhysRevLett.109.157002}.

Transport properties of double quantum dots with superconducting contacts
have been recently explored both experimentally
\cite{Hofstetter2009Oct,Herrmann_PhysRevLett.104.026801,Hofstetter_PhysRevLett.107.136801,Schindele_PhysRevLett.109.157002,
Das2012Nov,Fulop_PhysRevB.90.235412,Fulop_PhysRevLett.115.227003,Tan_PhysRevLett.114.096602,Sherman2017Mar}
and theoretically
\cite{Tanaka2010Feb,Eldridge2010Nov,Chevallier2011_PhysRevB.83.125421,Baranski2012May,Trocha2014_PhysRevB.89.245418, Sothmann2014Dec,Trocha2015Jun,Hussein2016_PhysRevB.94.235134,WrzesniewskiJPCM17}.
The theoretical investigations were however mostly devoted to
transport properties in a relatively weak coupling regime.
Various geometries of the system were considered, with the two dots attached to the leads forming
either serial \cite{Tanaka2010Feb}, T-shaped \cite{Baranski2012May} or CPS fork configurations
\cite{Sothmann2014Dec,Trocha2015Jun}.
In particular, the emergence of the triplet blockade and its influence on transport were analyzed,
as well as various Andreev bound states (ABS) splitting mechanisms \cite{Eldridge2010Nov, Trocha2015Jun}. 
Moreover, unconventional pairing \cite{Sothmann2014Dec} in the presence
of inhomogeneous magnetic field was predicted and the role
of the spin-orbit interaction on nonlocal entanglement was demonstrated \cite{Hussein2016_PhysRevB.94.235134}.
Other important aspects of transport in such systems,
such as the current and noise correlations \cite{Chevallier2011_PhysRevB.83.125421, WrzesniewskiJPCM17}
and spin-dependence of transport controlled by means of ferromagnetic contacts \cite{Trocha2014_PhysRevB.89.245418, Trocha2015Jun, WrzesniewskiJPCM17}, were also thoroughly discussed.

In this paper we extend those studies by focusing on the 
Andreev transport in the strong coupling regime,
where electronic correlations can give rise to the Kondo effect
\cite{Kondo_Prog.Theor.Phys32/1964,Hewson_book}.
When a spin one-half impurity is coupled to the conduction band of a metallic host,
for temperatures $T$ lower than the Kondo temperature $T_K$,
the conduction electrons screen the impurity's spin
and a delocalized singlet state is formed. Its emergence 
results in the formation of an additional peak at the Fermi energy
in the local density of states \cite{Hewson_book}. For single quantum dots,
in the case of spin $SU(2)$ Kondo effect,
this leads to an enhancement of the conductance
to its maximum value of $2e^2/h$ \cite{Goldhaber_Nature391/98,Cronenwett_Science281/1998}.
For double quantum dots, depending on the DQD occupation,
one can observe different types of the Kondo effect.
In particular, when both the spin and orbital degrees of freedom
are degenerate, an $SU(4)$ Kondo state is formed in the system \cite{Borda2003Jan,Keller2014Feb}.

When the leads are superconducting the situation
becomes much more interesting \cite{DeFranceschi2010,MartinRodero_AdinPhys2011,
Lee2012Oct,Pillet2013Jul,Lee2014Jan}.
First of all, for dot coupled to superconductor,
the occurrence of the Kondo phenomenon
is conditioned by the ratio of the Kondo temperature to the
superconducting energy gap $T_K/\Delta$,
and a quantum phase transition occurs as this ratio is varied
\cite{MartinRodero_AdinPhys2011,Bauer2007Nov,Maurand2013Jul,Franke2011May,Kim2013Feb}.
Furthermore, for two-terminal hybrid junctions
involving quantum dot and N and SC leads,
the Kondo state can be formed by screening the dot's spin
by the normal lead \cite{DeFranceschi2010,MartinRodero_AdinPhys2011},
while finite coupling to SC lead can result in an enhancement of the Kondo temperature \cite{Domanski2016Mar}.

From theoretical side, the accurate studies of transport properties
of nanostructures in nonperturbative regime require resorting to sophisticated
numerical methods. One of them is the density-matrix
numerical renormalization group (DM-NRG) method \cite{Wilson_Rev.Mod.Phys.47/1975,Legeza_DMNRGmanual},
which allows for obtaining results of very high accuracy
on the transport behavior of considered system \cite{Bulla_Rev.Mod.Phys.80/2008}.
In these considerations we employ DM-NRG to address the 
problem of the Kondo effect and Andreev transport 
in double quantum dot based Cooper pair splitters.
In particular, we study the DQD energy level dependence
of the local density of states as well as the Andreev transmission coefficient,
together with the splitting efficiency of the device.
We then focus on the two transport regimes
when the system in the absence of coupling to superconductor
exhibits either the $SU(2)$ or the $SU(4)$ Kondo effect,
and study the influence of superconducting pairing correlations on these two types of Kondo state.
We show that, contrary to single quantum dots \cite{Zitko2015Jan,Domanski2016Mar},
the $SU(2)$ Kondo state becomes quickly suppressed
by even small superconducting pairing potential.
On the other hand, the pairing correlations result
in a crossover from the $SU(4)$ to the $SU(2)$ Kondo effect.

The paper is organized as follows.
In Sec. II we present the model, Hamiltonian
and method used in calculations,
and describe the main quantities of interest.
Section III is devoted to numerical results and their discussion.
In Secs. III.A and III.B we analyze the DQD level dependence 
of the local density of states and the Andreev transmission coefficient,
together with splitting efficiency, respectively.
The $SU(2)$ [$SU(4)$] Kondo regime
is thoroughly discussed in Sec. III.C (Sec. III.D).
Finally, the conclusions can be found in Sec. IV.


\section{Theoretical description}
\label{sec:2}

\subsection{Model and parameters}


\begin{figure}[t]
\centering
\includegraphics[width=1\columnwidth]{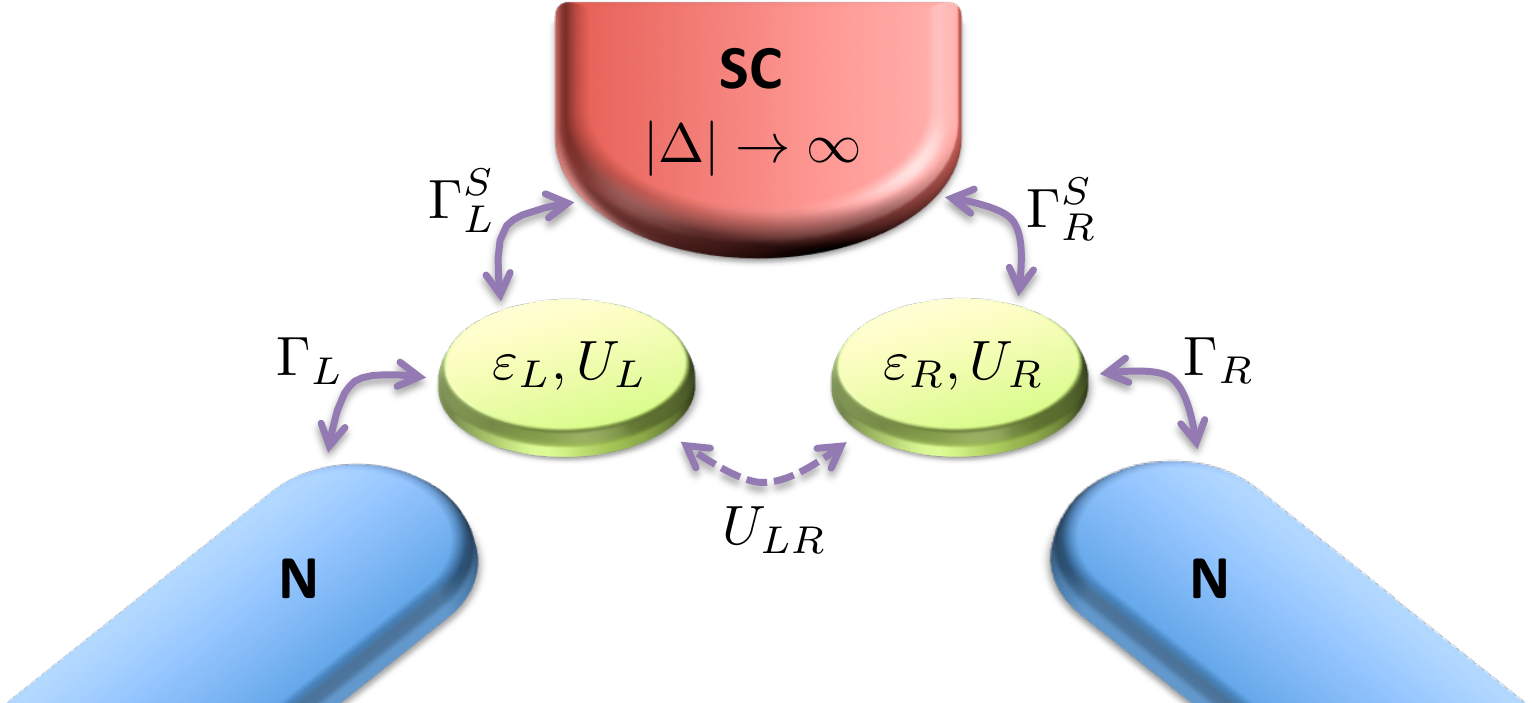}
\caption{\label{Fig:scheme}
Schematic of the considered system.
Two single-level quantum dots,
described by on-site energy $\e_j$
($j=L$ for left and $j=R$ for right dot)
and Coulomb correlations $U_j$
are coupled to a common $s$-wave superconductor (SC),
with coupling strength $\Gamma_j^S$,
and attached to two separate normal (N) electrodes,
with coupling strength $\Gamma_j$.
The two dots are coupled capacitively by $U_{LR}$.}
\end{figure}

The considered system consists of two single level quantum dots
attached to an $s$-wave superconductor (SC) and coupled
to two normal (N) electrodes, see \fig{Fig:scheme}.
The Hamiltonian of isolated double quantum dot has the form
\begin{eqnarray}\label{Eq:HDQD}
  H_{DQD}&=&\sum_{j\sigma} \varepsilon_j d_{j \sigma}^{\dagger} d_{j \sigma}
  +\sum_j U_j n_{j \uparrow}n_{j \downarrow} \nonumber\\
   &&+ \sum_{\sigma\sigma'} U_{LR} n_{L\sigma}n_{R\sigma'},
\end{eqnarray}
with $d_{j \sigma}^{\dagger}$ creating
a spin-$\sigma$ electron in dot $j$ of energy $\e_j$.
The on-dot Coulomb correlations are denoted by $U_j$,
with $n_{j \sigma} = d_{j \sigma}^{\dagger}d_{j \sigma}$,
while the inter-dot Coulomb interactions are described by $U_{LR}$.
The normal electrodes are modeled as free quasiparticles
by the Hamiltonian,
$H_{N} = \sum_{j\textbf{k}\sigma} \varepsilon_{j\textbf{k}}
c_{j \textbf{k}\sigma}^{\dagger}c_{j \textbf{k}\sigma}$,
where $c_{j \textbf{k}\sigma}^{\dagger}$ is the creation operator
for an electron with spin $\sigma$, wave number $\textbf{k}$ and
energy $\varepsilon_{j \textbf{k}}$ in the $j$th lead.
The BCS superconductor is modeled by
$H_{S}=\sum_{{\mathbf k}\sigma}  \xi_{{\mathbf k}}
     a^\dag_{{\mathbf k}\sigma}a_{{\mathbf k}\sigma}
     \!+\!
     \Delta \sum_{{\mathbf k}} \left( a_{{\mathbf
k}\downarrow}a_{{-\mathbf k}\uparrow} + {\rm h.c.}\right)$,
where $a^\dag_{{\mathbf k}\sigma}$ creates
an electron with momentum ${\mathbf{k}}$,
spin $\sigma$ and energy $\xi_{{\mathbf k}}$.
The superconducting order parameter,
assumed to be real, is denoted by $\Delta$.
The double dot is coupled to external leads by the
tunneling Hamiltonian
\begin{equation}\label{Eq:4}
H_T=\!\! \sum_{j\mathbf{k}\sigma}
    \!\big(V_{j\mathbf{k}} c^\dag_{j\mathbf{k}\sigma}d_{j \sigma}+
    V_{j\mathbf{k}}^S a^\dag_{\mathbf{k}\sigma}d_{j \sigma} +
   \rm h.c. \big),
\end{equation}
where the tunnel matrix elements between the dot $j$
and the normal lead $j$ (superconductor) are denoted
by $V_{j\mathbf{k}}$ ($V_{j\mathbf{k}}^S$).
Assuming momentum independent tunnel matrix elements,
the coupling between the dot $j$ and the corresponding
normal electrode is described by,
$\Gamma_j = \pi |V_j|^2\rho_j$,
where $\rho_j$ is the density of states of lead $j$.
On the other hand, the coupling between the dot $j$
and superconductor is given by, $\Gamma_j^S = \pi |V^S_j|^2\rho_S$,
with $\rho_S$ the density of states of superconductor in the normal state.

In our considerations we focus on the Andreev transport regime,
therefore, to exclude the normal tunneling processes,
in the following we take the limit of infinite superconducting energy gap.
In this limit the double dot coupled to superconductor
can be described by the effective Hamiltonian of the form~\cite{Rozhkov2000Sep,Eldridge2010Nov,Trocha2015Jun}
\begin{eqnarray}\label{Eq:HDQDeff}
  H_{DQD}^{\rm eff}=H_{DQD} -\sum_j \Gamma_j^S \left(d_{j\uparrow}^{\dagger}d_{j\downarrow}^{\dagger} + d_{j\downarrow}d_{j\uparrow}\right)
  \nonumber \\
  +\Gamma_{LR}^S \left(d_{L\uparrow}^{\dagger}d_{R\downarrow}^{\dagger} + d_{R\uparrow}^{\dagger}d_{L\downarrow}^{\dagger}+
  d_{R\downarrow}d_{L\uparrow} + d_{L\downarrow}d_{R\uparrow}
  \right).
\end{eqnarray}
Now, the proximity effect is included through pairing potential induced in the DQD,
where the first term, proportional to $\Gamma_j^S$, describes the direct Andreev reflection (DAR)
processes, while the last term, proportional to
$\Gamma_{LR}^S = \sqrt{\Gamma_{L}^S \Gamma_{R}^S}$,
corresponds to the crossed Andreev reflection
(CAR) processes. In DAR processes Cooper pairs are transferred
through one arm of the splitter. On the other hand, in CAR processes Cooper pair electrons
become split and each electron leaves the superconductor through different
junction with normal lead.

The effective double dot Hamiltonian is not diagonal
in the local basis defined by the states
$\ket{\chi_L\chi_R}=\ket{\chi_L}\ket{\chi_R}$,
in which the left (right) dot is in state $\ket{\chi_L}$ ($\ket{\chi_R}$),
with $\chi_j = 0,\sigma,d$, for empty, singly occupied and doubly occupied dot $j$.
Because the effective Hamiltonian commutes
with the total spin operator, 
$H_{DQD}^{\rm eff}$ has a block-diagonal form
in the corresponding spin quantum number. As we show in the Appendix,
the spin triplet space is quite trivial because it is not affected by
the superconducting correlations due to symmetry reasons.
In the spin doublet subspace we present a general solution to the eigen-problem.
However, in the singlet subspace it is in general
not possible to find simple analytical formulas for the eigenstates and eigenenergies,
therefore in this subspace we discuss the eigenspectrum
only in some limiting situations. The first one
is the particle-hole symmetry point of the model,
$\e = -U/2-U_{LR}$, and the second one
is the fully symmetric $SU(4)$ Kondo regime,
$\e = -U_{LR}/2$ with $U_{LR}=U$.
The analytical formulas presented in the Appendix
will be crucial to understand the complex behavior
of the system in the considered transport regimes.
Moreover, the eigenenergies will help to 
relate the position of peaks observed in transport quantities
to energies of Andreev bound states (ABS),
which can be inferred from excitation
energies between the corresponding molecular states
of the double quantum dot proximized by SC lead.

In our analysis we assume that the system is symmetric, i.e. we set
$\Gamma_L = \Gamma_R \equiv \Gamma$
and
$\Gamma_L^S = \Gamma_R^S \equiv \Gamma_S$.
For the two quantum dots we also assume
$U_L = U_R \equiv U$ and $\e_L = \e_R \equiv \e$.
To perform the calculations, we set $U\equiv 1$
and take $U_{LR} = U/2$ and $\Gamma = U/20$.
We note that since both the couplings and the position of the DQD levels
can be tuned individually by applied gate voltages
\cite{Keller2014Feb},
the chosen set of parameter is of relevance for current and future experiments.
We also notice that a weak left-right asymmetry
would induce rather quantitative changes to the results we present and discuss in the following,
while qualitatively we expect our predictions to be relevant.
However, the assumption of the superconducting energy gap being
the largest energy scale in the problem needs to be treated with a certain care.
While this assumption allows us to focus exclusively on the behavior of
Andreev reflection processes, and for that reason it was
adapted in many previous theoretical works
\cite{Yoichi2007Jun,Meng2009Jun,Futterer2009Feb,Tanaka2010Feb,Eldridge2010Nov,Sothmann2010Sep,Baranski2011Nov,
Futterer2013Jan,Wojcik2014Apr,Sothmann2014Dec,Weymann2014Mar,Weymann2015Dec,
Trocha2015Jun,Chirla2016Jul,Glodzik2017Mar},
from experimental point of view,
the condition $\Delta > U$ does not need to be fulfilled
in any Cooper pair splitting device.
Nevertheless, there are superconductors, in which the gap is 
of the order of a couple of meV
\cite{Heinrich2013Dec}, consequently,
experimental realizations of splitters
with large $\Delta$ should be possible.

\subsection{Quantities of interest and method}

The main quantity of interest is the transmission
coefficient for Andreev reflection processes, $\TA $, which can be written as
\beq
\TA  =  \TAD + \TAC,
\eeq
where the first term describes the transmission due to DAR processes,
which is explicitly given by
\beq
\TAD = 4 \sum_{j\sigma} \Gamma_{j}^2  |\zub{d_{j\sigma}}{d_{j\bar{\sigma}}}^r_\omega|^2 \,,
\eeq
while the last term denotes the transmission coefficient due to CAR processes and is described by
\beq
\TAC = 4 \Gamma_{L}\Gamma_R \! \sum_\sigma \! \big[ |\zub{d_{L\sigma}}{d_{R\bar{\sigma}}}^r_\omega|^2 +
 |\zub{d_{R\sigma}}{d_{L\bar{\sigma}}}^r_\omega|^2 \big] .
\eeq
Here, $\zub{A}{B}_\omega^r$ is the Fourier transform of the retarded Green's function,
$\zub{A}{B}_t^r = -i\Theta(t) \expect{\{ A (t),B(0) \}}$.
The DAR and CAR transmission coefficients can be used
to define the Cooper pair splitting efficiency of the device as
\beq \label{eq:eta}
\eta = \frac{\TAC}{\TAC+\TAD}.
\eeq
When $\eta \to 1$, transport is exclusively due to CAR processes,
which means that each Cooper pair leaving superconductor
becomes split into two separate leads.
On the other hand, if only DAR processes are responsible for Andreev transport, $\eta \to 0$.

With the Andreev transmission coefficient, it is possible 
to determine the Andreev current flowing between the superconductor
and the normal leads \cite{Trocha2014_PhysRevB.89.245418}
\beq \label{Eq:I}
  I_A(V) = \frac{e}{h} \int \!\! d\omega \left[ f(\omega-eV)-f(\omega+eV) \right] \TA ,
\eeq
where $f(\omega)$ denotes the Fermi-Dirac distribution function
and it is assumed that the chemical potential of the left and right lead
is equal to $eV$, while the superconductor is grounded.
From the above formula it is easy to find the Andreev differential conductance,
which in the limit of vanishing temperature can be approximated by
\beq \label{Eq:G}
  G_A(V) \approx \frac{e^2}{h} \left[ T_A(\omega=eV) + T_A(\omega=-eV) \right].
\eeq
Consequently, the measurement of differential conductance
allows one to probe the energy dependence of the Andreev transmission coefficient.

Another interesting quantity is the local density of states,
which is given by the total normalized spectral function
\beq
\A = \sum_{ij}\A_{ij} = - \sum_{ij} \sqrt{\Gamma_i \Gamma_j} \;{\rm Im} \zub{d_{i\sigma}}{d_{j\sigma}^\dag}_{\omega}^r.
\eeq
Thus, $\A_{i}\equiv \A_{ii}$ corresponds to the local density of states
of one of the quantum dots, while $\A_{ij}$ describes
the cross-correlations between the two quantum dots
generated by proximity-induced inter-dot pairing potential $\Gamma_{LR}^S$.
Because we consider symmetric situation,
$\A_L = \A_R$, and $\A_{LR }= \A_{RL}$.

To determine the relevant correlation functions we use the density-matrix
numerical renormalization group method \cite{Wilson_Rev.Mod.Phys.47/1975,
Bulla_Rev.Mod.Phys.80/2008,Legeza_DMNRGmanual}.
This nonperturbative method allows for obtaining 
very accurate results on the static and dynamic properties of the system.
In NRG, the initial Hamiltonian is transformed to an NRG Hamiltonian, in which
the leads are modeled as tight-binding chains with appriopriate hopping intergrals
\cite{Wilson_Rev.Mod.Phys.47/1975}.
The calculations are performed in an iterative fashion by keeping
an assumed number $N_K$ of the lowest-energy eigenstates.
Here, we exploited the full spin symmetry of the system and kept at least $N_K=2000$
states per iteration. The imaginary parts of the Green's functions
were determined from discrete NRG data
by performing appropriate broadening \cite{Freyn2009Mar} and averaging over $N_z=2$
shifted discretization meshes \cite{OliveiraPhysRevB.49.11986}.
The real parts of the Green's functions
were obtained from the Kramers-Kronig relation.

\subsection{Stability diagram and transport regimes}

\begin{figure}[t]
\centering
\includegraphics[width=0.85\columnwidth]{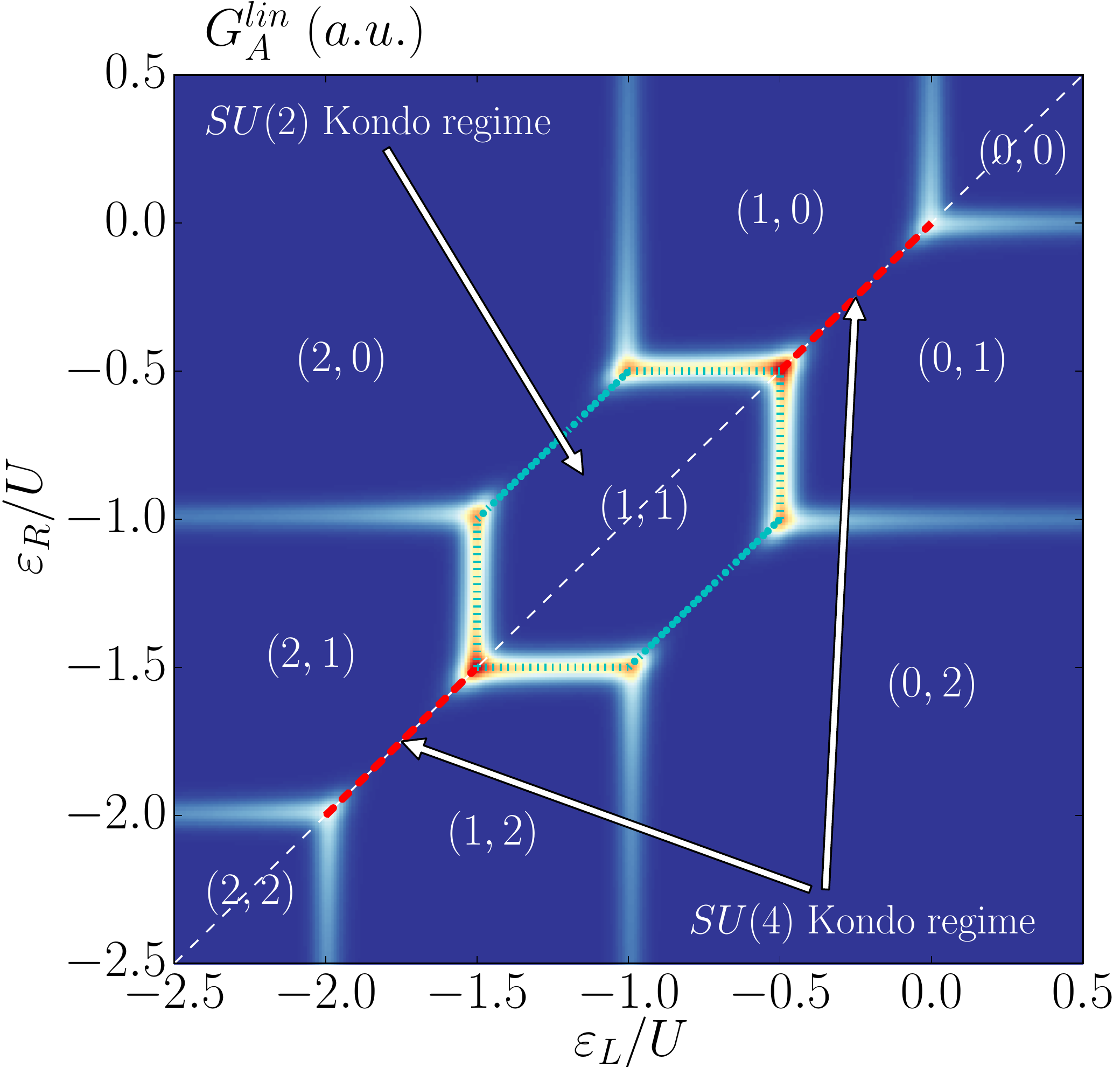}
\caption{\label{Fig:stab}
The linear Andreev conductance $G_A^{lin}$ calculated
as a function of the position of each dot level,
$\e_L$ and $\e_R$, using the rate equations.
The parameters are: $U=1$, $U_{LR}=U/2$, $\Gamma=U/100$,
and $\Gamma_S=U/10$.
The numbers in brackets indicate the average occupation
of each dot, $(\expect{n_L},\expect{n_R})$,
with $n_j = \sum_\sigma n_{j\sigma}$.}
\end{figure}

The linear Andreev conductance plotted as a function
of the position of each dot level assuming a weak coupling
between the double dot and normal leads is shown in \fig{Fig:stab}.
The numbers in brackets indicate approximate expectation values of the occupation number 
of each dot, $(\expect{n_L},\expect{n_R})$,
with $n_j = \sum_\sigma n_{j\sigma}$.
The conductance was calculated using the rate equations
within the sequential tunneling approximation \cite{Trocha2015Jun}.
We note that although this method is not suitable for capturing the 
correlation effects studied here, it 
allows us to indicate the considered transport regimes
in the phase diagram of the device.
In this paper we in particular focus on the 
symmetric case, $\e_L = \e_R \equiv \e$,
a cross-section of \fig{Fig:stab} marked with a dashed line.
By sweeping $\e$, which can be experimentally done with gate voltages
\cite{Keller2014Feb},
the device can be tuned from the empty or fully-occupied orbital regime 
to the $SU(4)$ and $SU(2)$ Kondo regimes, respectively.\footnote{
Note that the $SU(2)$ and $SU(4)$ Kondo 
regimes can be greatly modified by finite coupling to superconductor,
such that the Kondo effect can even become fully suppressed.
Therefore, referring to the appropriate Kondo regime
should be considered as a guide to estimate the corresponding parameter space
in the phase diagram of the device in the limit of weak coupling to superconductor.}
The $SU(4)$ Kondo regime is marked with a thick dashed line,
while the $SU(2)$ Kondo regime is surrounded
by dotted lines in \fig{Fig:stab}.
These transport regimes will be studied in detail
in the next sections, and the influence 
of the proximity-induced pairing potential
on the corresponding Kondo states will be thoroughly analyzed.

\section{Results and discussion}

In this section we present and discuss the main results
on the local density of states and the Andreev transmission coefficient.
We will first study the general gate voltage dependence
of transport characteristics assuming  $\e_L = \e_R \equiv \e$,
i.e. along the dashed line marked in \fig{Fig:stab}.
Then, we shall focus on some more relevant transport regions,
including the $SU(2)$ and $SU(4)$ Kondo regimes.

\subsection{Local density of states}

\begin{figure}[t]
\centering
\includegraphics[width=0.97\columnwidth]{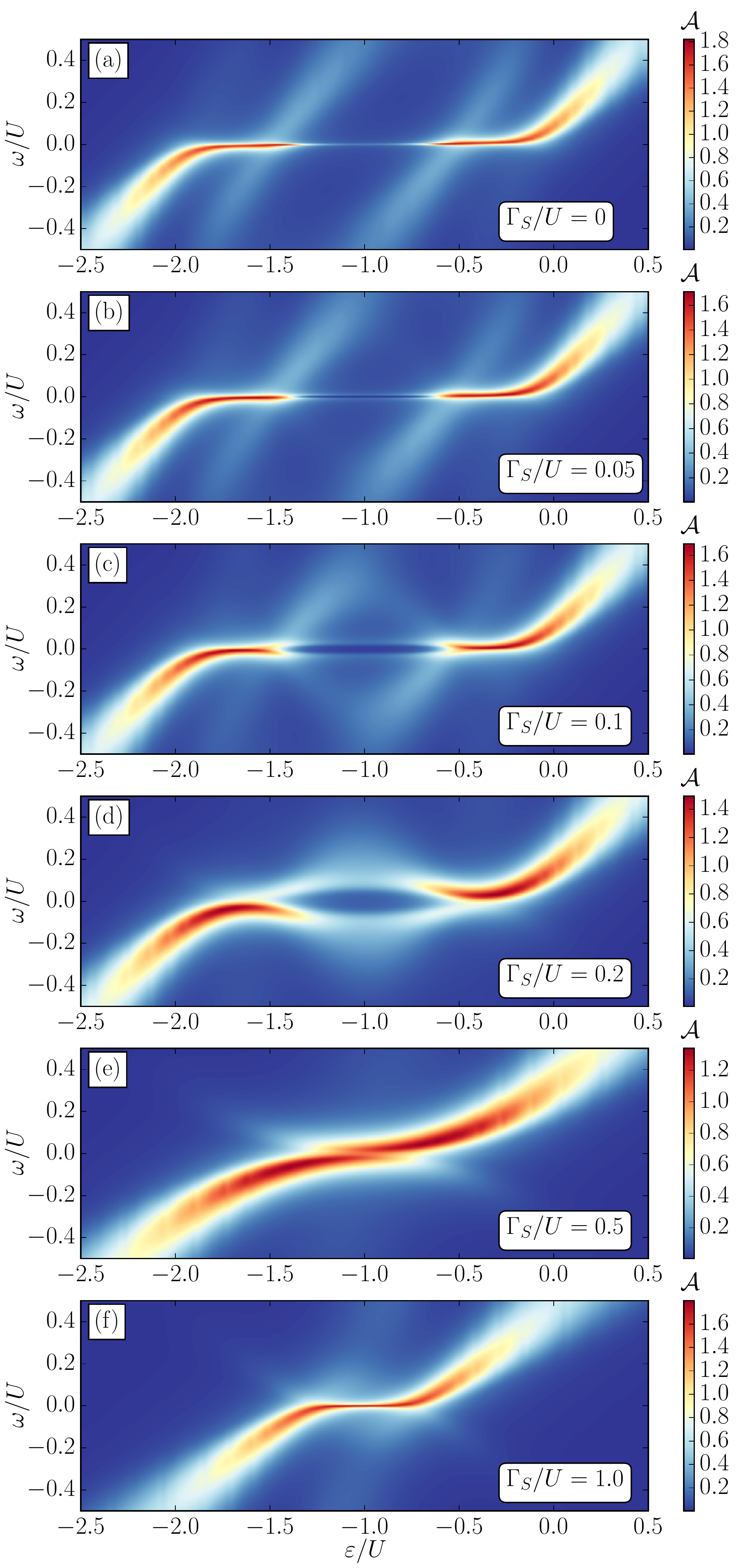}
\caption{\label{Fig:AGammaS}
The total normalized spectral function $\A$
of DQD-based Cooper pair splitter 
plotted as a function of energy $\omega$
and double dot level position, $\e_L = \e_R \equiv \e$,
calculated for different values
of coupling to superconductor $\Gamma_S$, as indicated.
The parameters are:
$U=1$, $U_{LR}=U/2$, $\Gamma=U/20$, and $T=0$.}
\end{figure}

The normalized spectral function plotted
as a function of energy $\omega$ and DQD level position
$\e_L = \e_R \equiv \e$ is shown in \fig{Fig:AGammaS}.
This figure is calculated for different values of 
the coupling to superconductor, as indicated,
and it demonstrates the evolution of local density of states
with increasing $\Gamma_S$.
When $\Gamma_S=0$, one observes the transport behavior
typical for a double quantum dot system \cite{Wiel_Rev.Mod.Phys.75/2002},
see \fig{Fig:AGammaS}(a).
When the position of the DQD energy levels is lowered,
the DQD becomes consecutively occupied with electrons.
For $\e \gtrsim 0$ ($\e\lesssim -U-2U_{LR}$),
the DQD is empty (fully occupied). When $-U_{LR} \lesssim \e \lesssim 0$
($-U-2U_{LR} \lesssim \e \lesssim -U-U_{LR}$),
the double dot is singly occupied (occupied with three electrons),
while for $-U-U_{LR} \lesssim \e \lesssim -U_{LR} $,
the DQD is occupied by two electrons, each located on different quantum dot.
The above energies also specify when the charge on the DQD changes
and the local density of states exhibits a resonance.
In between those resonant energies, the system's spectral function
exhibits an enhancement due to the Kondo effect.
In the odd occupation regime, i.e. when DQD hosts either one or three electrons,
the system exhibits the $SU(4)$ Kondo effect
resulting from orbital and spin degeneracies
\cite{Borda2003Jan,Keller2014Feb}.
One can estimate the $SU(4)$ Kondo temperature, $\TKf$,
from the halfwidth at half maximum (HWHM) of the Kondo peak in the total spectral function for $\e = -U_{LR}/2$,
which for assumed parameters yields, $\TKf/U \approx 0.017$.
On the other hand, when the DQD is occupied by two electrons,
each dot exhibits the spin $SU(2)$ Kondo resonance
\cite{Hewson_book,Goldhaber_Nature391/98}.
The corresponding Kondo temperature, $\TKt$, estimated from
HWHM of the Kondo resonance in the spectral function
for $\e = -U/2-U_{LR}$, is equal to, $\TKt/U\approx 10^{-4}$.
Note that for the parameters assumed in calculations $\TKt \ll \TKf$.
This is why in \fig{Fig:AGammaS}(a)
the $SU(2)$ Kondo peak is much less 
pronounced as compared to the $SU(4)$ Kondo resonance.

\begin{figure}[t]
\centering
\includegraphics[width=1\columnwidth]{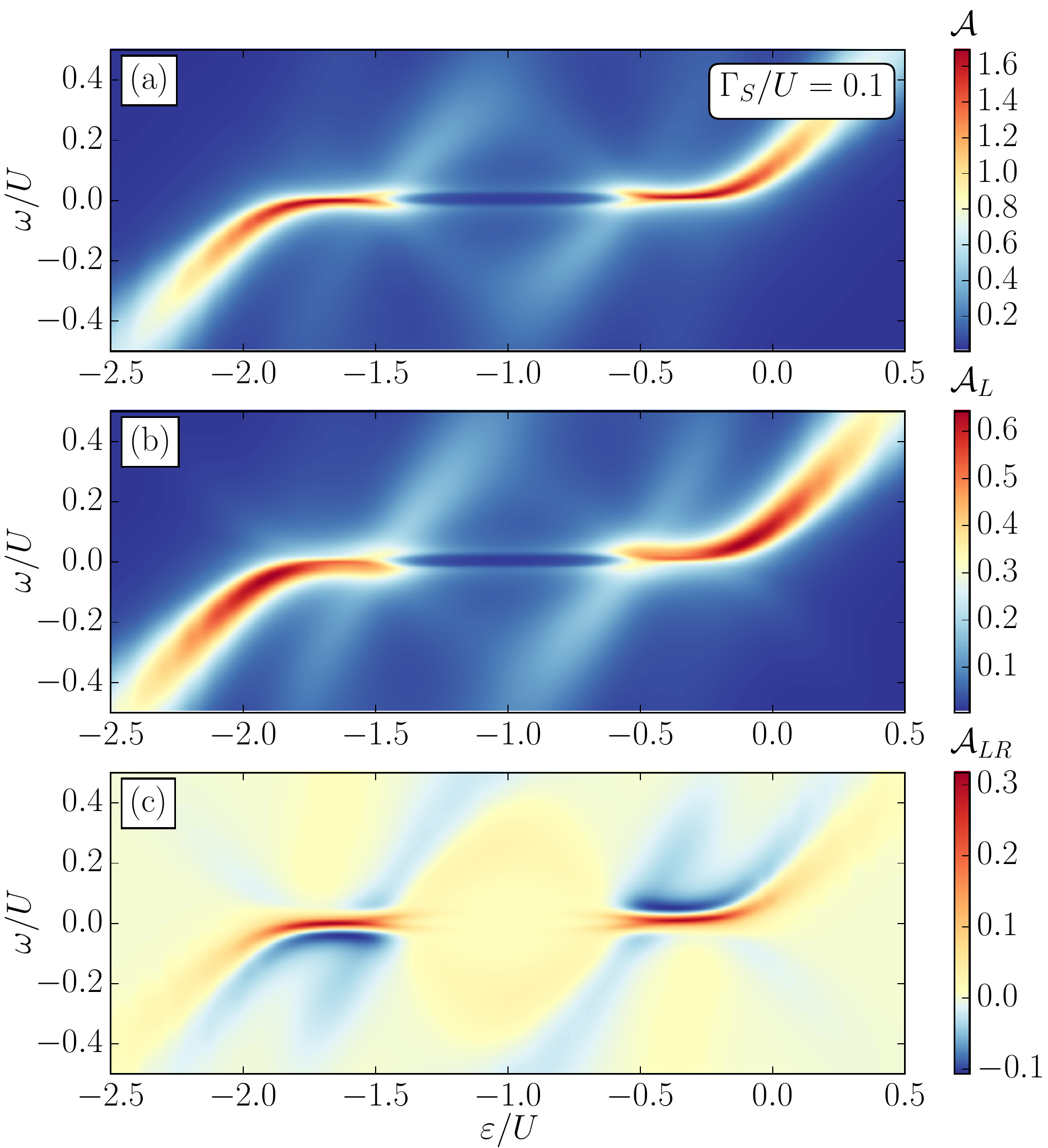}
\caption{\label{Fig:A}
The normalized spectral function: (a) $\A$, (b) $\A_{L}$ and (c) $\A_{LR}$,
plotted versus energy $\omega$
and double dot energy level position $\e$.
The parameters are the same as in \fig{Fig:AGammaS} with 
$\Gamma_S=U/10$.
}
\end{figure}

When the coupling to superconductor becomes finite,
the behavior of the spectral function starts changing.
First, one observes the suppression and splitting of the 
Kondo resonance in the doubly occupied transport regime,
see Figs. \ref{Fig:AGammaS}(b)-(d).
This splitting increases with $\Gamma_S$, however,
when $\Gamma_S \gtrsim U/2$, a single resonance starts forming,
see Figs. \ref{Fig:AGammaS}(e) and (f).
This resonance is again due to the Kondo effect,
since for $\Gamma_S \gtrsim U/2$, the doublet state
becomes the ground state of the system.
On the other hand, the $SU(4)$ Kondo resonance looks much less affected,
at least for small values of coupling to superconductor.
This is, however, not entirely true, since with increasing
$\Gamma_S$, the $SU(4)$ Kondo resonance merges with
resonance resulting from the formation of Andreev bound states.
A thorough discussion of the influence of strength of coupling to superconductor
on the corresponding Kondo resonances will be presented in next sections.

Let us now analyze the behavior of separate contributions,
$\A_L$ and $\A_{LR}$, to the total spectral function $\A$.
Their energy and DQD energy level dependence 
is shown in \fig{Fig:A} for $\Gamma_S = U/10$.
At first sight, one can notice that the qualitative behavior of $\A$
is mainly determined by the spectral function of 
single quantum dot $\A_L$. 
For the considered value of $\Gamma_S$,
$\A_L$ exhibits a pronounced split Kondo resonance
for $-U-U_{LR}\lesssim \e \lesssim -U_{LR}$
and the $SU(4)$ Kondo resonance 
when $-U_{LR}\lesssim \e \lesssim 0$
($-U-2U_{LR}\lesssim \e \lesssim -U-U_{LR}$),
similarly to the total spectral function,
cf. Figs. \ref{Fig:A}(a) and (b).

On the other hand, 
the off-diagonal spectral function,
which accounts for the cross-correlations between transport processes
through the two dots, behaves in a clearly different manner.
First of all, we note that finite value of $\A_{LR}$ results solely from 
proximity-induced inter-dot pairing,
and it vanishes if CAR processes are not allowed in the system.
One can see that $\A_{LR}$ takes considerable values
for energies corresponding to resonances in $\A$,
cf. Figs. \ref{Fig:A}(a) and (c).
Moreover, if on one side of the resonance 
$\A_{LR}$ is positive, on the other side it changes sign.
This effect is most pronounced for 
$-U_{LR}\lesssim \e \lesssim 0$
($-U-2U_{LR}\lesssim \e \lesssim -U-U_{LR}$), 
i.e. when DQD hosts an odd number of electrons, see \fig{Fig:A}(c). 
Positive sign of $\A_{LR}$
can be associated with processes that occur in the same
direction through both normal junctions, while
negative sign of $\A_{LR}$ indicates
that the two processes are anti-correlated
\cite{WrzesniewskiJPCM17}.

\subsection{Andreev transmission and splitting efficiency}

\begin{figure}[t]
\centering
\includegraphics[width=1\columnwidth]{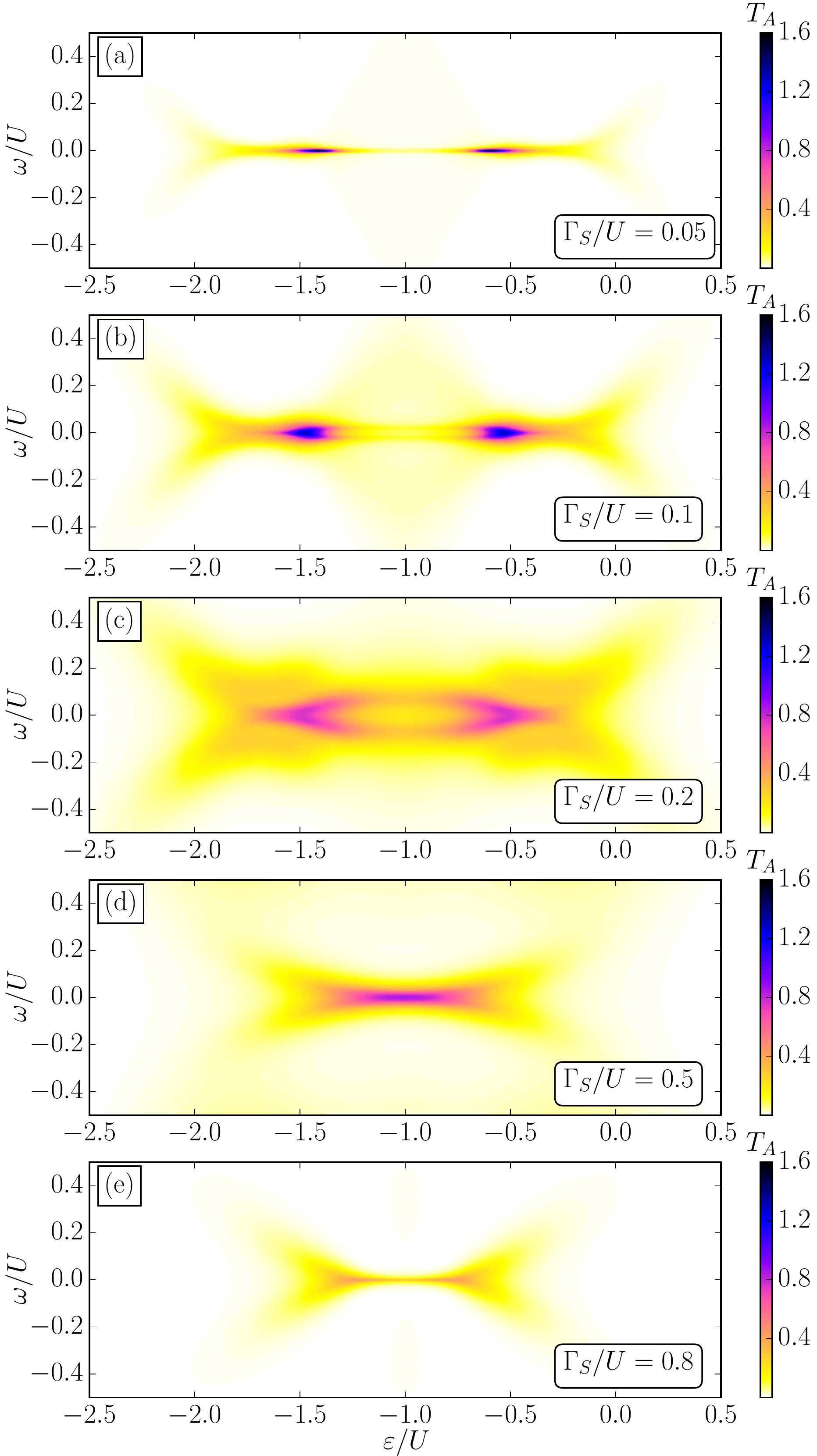}
\caption{\label{Fig:TAGammaS}
The Andreev transmission coefficient 
plotted versus energy $\omega$
and double dot energy level position $\e$,
and calculated for different values of $\Gamma_S$, as indicated.
The other parameters are the same as in \fig{Fig:AGammaS}.}
\end{figure}

\begin{figure}[t]
\centering
\includegraphics[width=1\columnwidth]{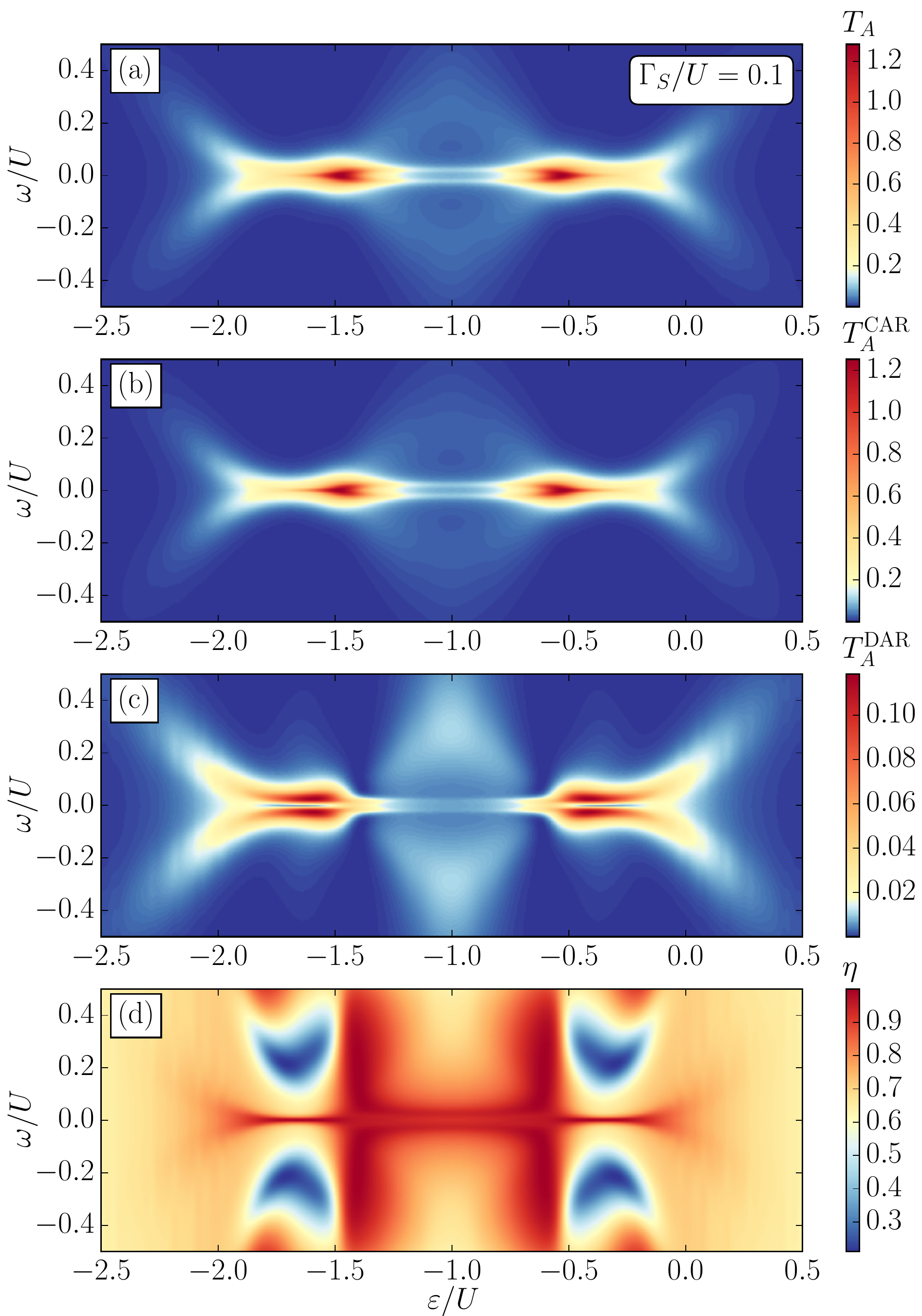}
\caption{\label{Fig:TA}
(a) The total Andreev transmission coefficient
and its contributions due to (b) CAR and (c) DAR
processes, as well as (d) Cooper pair splitting efficiency $\eta$
plotted as function of energy $\omega$
and double dot energy level position $ \e$.
The parameters are the same as in \fig{Fig:AGammaS} with 
$\Gamma_S=U/10$.
}
\end{figure}

The energy and DQD level dependence of the Andreev transmission coefficient
calculated for different values of coupling to superconductor
is presented in \fig{Fig:TAGammaS}.
When the coupling $\Gamma_S$ is relatively small,
one can see that $\TA$ becomes finite in the low-energy
regime and it is considerably enhanced 
for $\e \approx -U_{LR}$
and $\e \approx -U-U_{LR}$, see \fig{Fig:TAGammaS}(a).
The area when the maximum occurs grows
with increasing $\Gamma_S$ and, at the same time,
the maximum value slightly decreases.
Moreover, for $\Gamma_S=U/5$,
$\TA$ becomes finite in almost the whole energy range considered in the figure,
with maximum values occurring still 
for $\e \approx -U_{LR}$ and $\e \approx -U-U_{LR}$, see \fig{Fig:TAGammaS}(c).
Note that $\TA$ exhibits a similar split structure
as that visible in the local density of states,
cf. Figs. \ref{Fig:AGammaS}(d) and \ref{Fig:TAGammaS}(c).
Further increase of the coupling strength
results in a decrease of the size of the Coulomb blockade regime,
which is seen as merging of the two maxima
at the particle-hole symmetry point $\e = -U/2-U_{LR}$
[\fig{Fig:TAGammaS}(d)]. For even larger $\Gamma_S$
the transmission coefficient drops and the energy range
where $\TA$ is enhanced shrinks, see 
\fig{Fig:TAGammaS}(e).

The different contributions to the transmission coefficient
coming from DAR and CAR processes are presented 
in \fig{Fig:TA} for $\Gamma_S=U/10$.
The first general observation is that the total
Andreev transmission is mainly determined by crossed Andreev reflection
processes. This can be expected because the 
rate of direct Andreev reflection is conditioned by the value of on-site Coulomb correlations,
while the rate of CAR processes depends on the inter-dot correlations.
Because $U_{LR}<U$, as in typical experimental realizations \cite{Hofstetter2009Oct},
one finds more CAR processes compared to DAR ones.
This is in fact a very desired situation for Cooper pair splitting experiments,
in which one would like to suppress DAR processes and maximize CAR ones.

\begin{figure}[t]
\centering
\includegraphics[width=1\columnwidth]{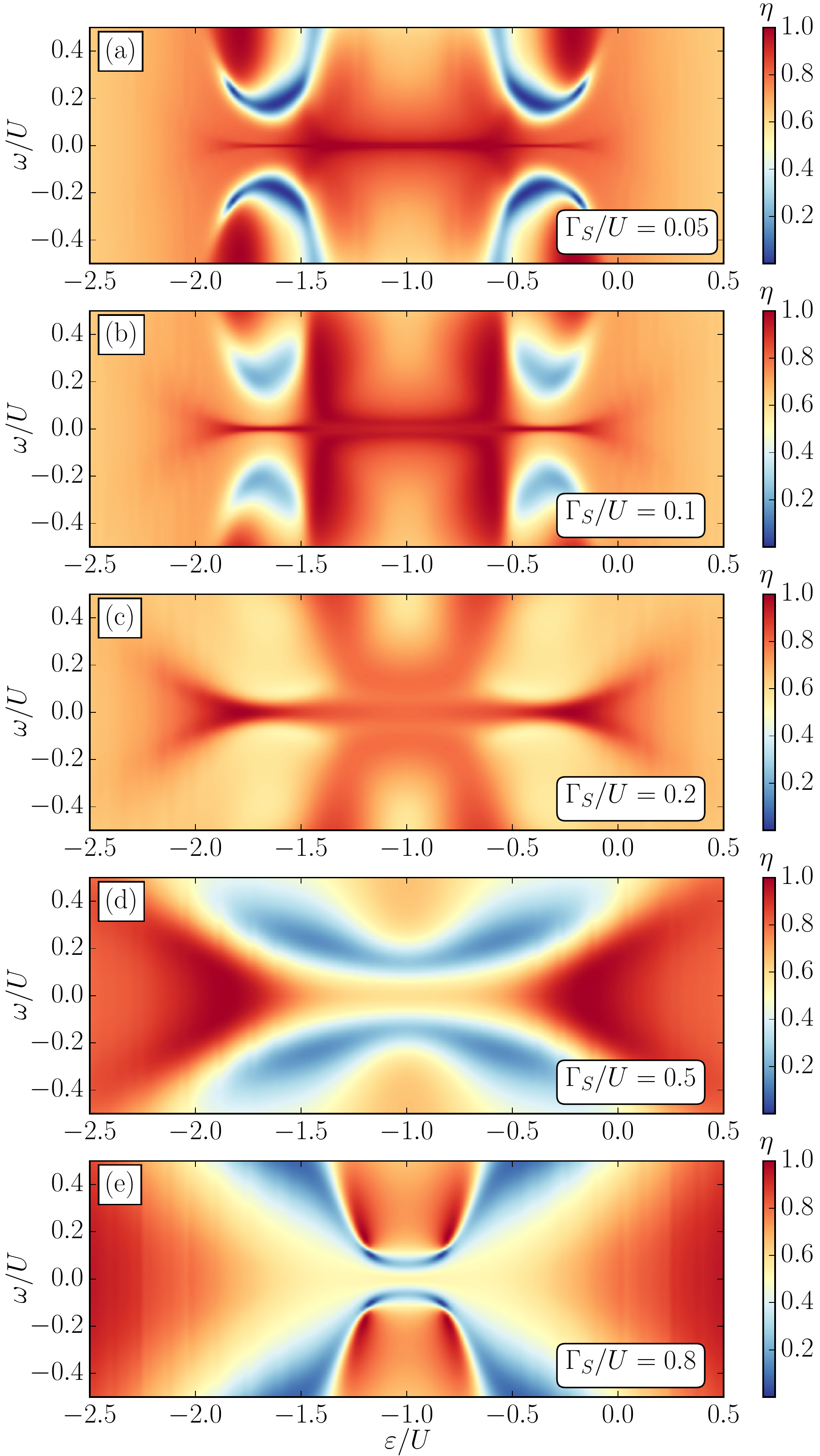}
\caption{\label{Fig:eta}
The splitting efficiency $\eta$ calculated for different values
of coupling strength to superconducting lead, as indicated,
and for parameters the same as in \fig{Fig:AGammaS}.}
\end{figure}

From the application point of view, it is thus interesting
to analyze the Cooper pair splitting efficiency $\eta$ of the device.
This is presented in \fig{Fig:TA}(d).
One can see that the splitting efficiency, depending
on DQD energy level position $\e$ and energy $\omega$, takes values
ranging from very low ($\eta \approx 0.2$) to its maximum value of $\eta = 1$.
We recall that for $\eta = 1$ transport is exclusively due to CAR processes,
while for $\eta = 0$, only DAR processes contribute to Andreev conductance,
cf. \eq{eq:eta}.
Clearly, large splitting efficiency is observed
at low energies and for $-U-2U_{LR} \lesssim \e \lesssim 0$,
see \fig{Fig:TA}(d). Moreover, a region of enhanced $\eta$
is present in the Coulomb blockade regime with two electrons.
Then, mainly CAR processes are responsible for Andreev transport.
Note also that there are transport regimes where
the splitting efficiency is rather poor
and mainly DAR processes are responsible for transport,
see the transport regime with odd number of electrons for elevated energies $|\omega|$
in \fig{Fig:TA}(d).

The splitting efficiency greatly depends on the strength of coupling to superconductor.
This dependence is explicitly demonstrated in \fig{Fig:eta},
which shows the energy and DQD level dependence of $\eta$
calculated for different values of $\Gamma_S$ corresponding to those
considered in \fig{Fig:TAGammaS}. 
In this figure one can identify optimal parameters,
for which the process of Cooper pair splitting is most efficient in the
considered transport regime.

Finally, we would like to emphasize that the splitting efficiency
also strongly depends on the ratio of interdot and intradot Coulomb correlations
$U/U_{LR}$. In typical experimental realizations, $U\gg U_{LR}$,
which is desired to enhance CAR processes and suppress DAR ones,
obtaining thus large values of $\eta$. The splitting efficiency however
generally decreases when the ratio of $U/U_{LR}$ becomes smaller.
In particular, the amount of DAR and CAR processes becomes equal
when $U=U_{LR}$, such that $\eta = 1/2$ in the whole parameter space.

\subsection{The SU(2) Kondo regime}

\begin{figure}[t]
\centering
\includegraphics[width=1\columnwidth]{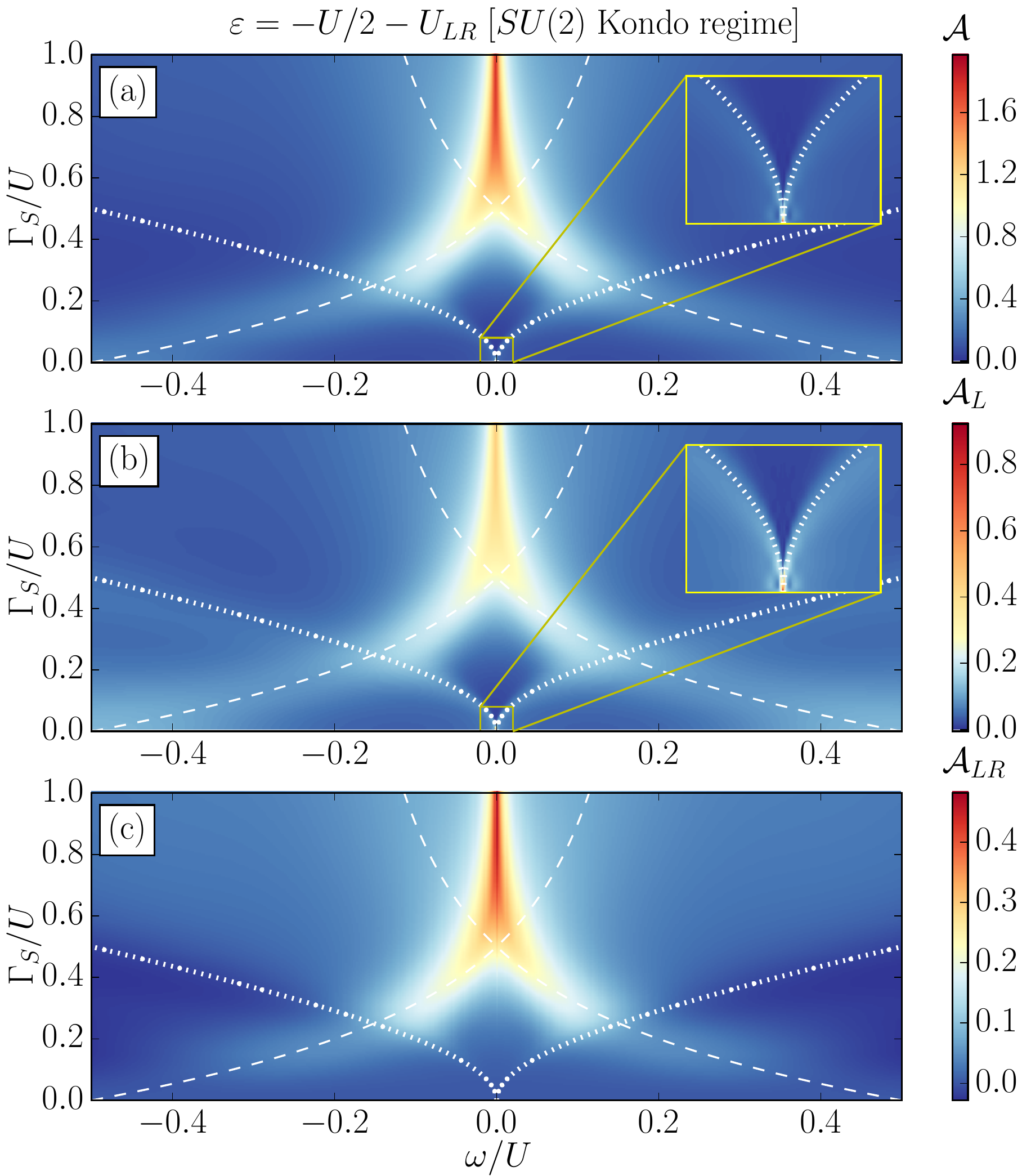}
\caption{\label{Fig:ASU2}
The energy dependence of (a) the total normalized spectral function $\A$
and its contributions: (b) $\A_L$ and (c) $\A_{LR}$
calculated as a function of the coupling to superconductor $\Gamma_S$
and for $\e =-U/2-U_{LR}$. 
The insets show the zoom into the suppression of the $SU(2)$ Kondo resonance
with increasing $\Gamma_S$.
The dashed lines indicate the energies of the Andreev bound states,
cf. \eq{eq:ABSsu2}, while the dotted lines present
the excitation energies between corresponding singlet and triplet states,
cf. \eq{Eq:sidepeaks}.
The other parameters are as in \fig{Fig:AGammaS}.
}
\end{figure}

We now focus in greater detail on the $SU(2)$ Kondo regime,
where for $\Gamma_S =0$ the DQD is occupied by two electrons,
each on different quantum dot, see \fig{Fig:stab}.
To simplify the discussion, we consider the particle-hole symmetry point
of the model, $\e = -U/2 - U_{LR}$. Nevertheless,
the conclusions drawn here shall apply to the whole two-electron Coulomb blockade regime
where the spin $SU(2)$ Kondo effect can develop.

The total normalized spectral function
in the $SU(2)$ Kondo regime, together with its contributions $\A_L$ and $\A_{LR}$,
calculated as a function of $\Gamma_S$ for $\e = -U/2 - U_{LR}$,
is shown in \fig{Fig:ASU2}.
The dashed lines indicate the energies of the Andreev bound states,
while the insets present the zooms into the low-energy
behavior of the spectral function, where the suppression
of the Kondo resonance with increasing $\Gamma_S$ is clearly visible.
The general behavior is as follows:
Finite coupling to superconductor results in the 
splitting and suppression of the Kondo resonance,
which however, emerges again for $\Gamma_S\approx U/2$.
In fact, for this value of $\Gamma_S$, the system exhibit a phase transition
and the ground state changes from spin singlet to spin doublet.
Consequently, the Kondo resonance develops
once $\Gamma_S\gtrsim U/2$, see \fig{Fig:ASU2}.

Let us shed more light on the system's behavior by using some analytical arguments.
For the particle-hole symmetry point, it is easy to find
the eigenspectrum of the effective Hamiltonian (\ref{Eq:HDQDeff}).
We will consider the lowest-energy singlet ($\ket{S}$), doublet ($\ket{D_\s}$)
and triplet ($\ket{T_\d}$) states. The first two states have the following explicit form
\begin{eqnarray} \label{Eq:en}
\ket{S} &=& \alpha \left(\ket{dd}-\ket{00}\right) - \beta \left(\ket{\up\down}-\ket{\down\up}\right), \nonumber\\
\ket{D_\sigma} &=& \frac{1}{2} \left(\ket{\sigma0} + \ket{0\sigma} + \ket{\sigma d} + \ket{d\sigma} \right), \nonumber
\end{eqnarray}
where the coefficients are given by, $\alpha = \sqrt{(\gamma-U-U_{LR})/(4\gamma)}$,
$\beta =  2\Gamma_S / \sqrt{\gamma(\gamma-U-U_{LR})}$,
and $\gamma = \sqrt{(U+U_{LR})^2 + 16\Gamma_S^2}$.
Note that these states correspond to the states $\ket{D_\s^2}$
and $\ket{S_4}$ presented in the Appendix.
The triplet state is three-fold degenerate with components:
$\ket{T_{+}}=\ket{\!\up\up}$,
$\ket{T_{-}}=\ket{\!\down\down}$,
and 
$\ket{T_{0}}=(\ket{\!\up\down} + \ket{\!\down\up})/\sqrt{2}$.
The energies of the above states are given by
\begin{eqnarray} \label{Eq:en}
E_S &=& -\frac{1}{2}\left[ U+U_{LR} + \sqrt{(U+U_{LR})^2 + 16 \Gamma_S^2}\right], \nonumber\\
E_D &=& -\frac{1}{2} \left(U+2U_{LR} + 4\Gamma_S \right), \\
E_T &=& -U-U_{LR}, \nonumber
\end{eqnarray}
respectively. Note that the energy of the triplet state does not depend on $\Gamma_S$.
This is to be expected since the triplet state does not match the symmetry of the 
$s$-wave superconductor.
The excitation energies between singlet and doublet
states define the relevant ABS's energies
\begin{equation} \label{eq:ABSsu2}
E_{ABS} = \pm\frac{U_{LR}}{2} \pm 2\Gamma_S \mp \frac{1}{2}\sqrt{(U+U_{LR})^2 + 16 \Gamma_S^2} ,
\end{equation}
which are marked with dashed lines in \fig{Fig:ASU2}.

In the case of $\Gamma_S = 0$, the singlet and triplet state are degenerate 
and the system exhibits the $SU(2)$ Kondo effect on each quantum dot,
see the insets in Figs. \ref{Fig:ASU2}(a) and \ref{Fig:ASU2}(b).
However, when $\Gamma_S$ becomes finite, 
the induced inter-dot pairing relevant for crossed Andreev reflection
results in the singlet-triplet splitting and 
causes the singlet state $\ket{S}$ to be the ground state of the system.
Because of that, the Kondo resonance gets very quickly suppressed 
when $\Gamma_S$ increases and only split Kondo peaks are visible,
see the insets in \fig{Fig:ASU2}.
The position of the split Kondo peaks is determined
by the excitation energy between the singlet and triplet states,
such that the peaks occur for
\beq \label{Eq:sidepeaks}
\omega \approx \pm \frac{1}{2} 
\left[U+U_{LR}-\sqrt{(U+U_{LR})^2 + 16 \Gamma_S^2}\right].
\eeq
Thus, for small values of $\Gamma_S$, the position of side
peaks depends in a parabolic way on the coupling to superconductor,
$\omega \approx \pm 4\Gamma_S^2 / (U+U_{LR}) $.
This parabolic dependence can be seen in 
Figs. \ref{Fig:ASU2}(a) and \ref{Fig:ASU2}(b)
and the corresponding insets.

\begin{figure}[t]
\centering
\includegraphics[width=0.95\columnwidth]{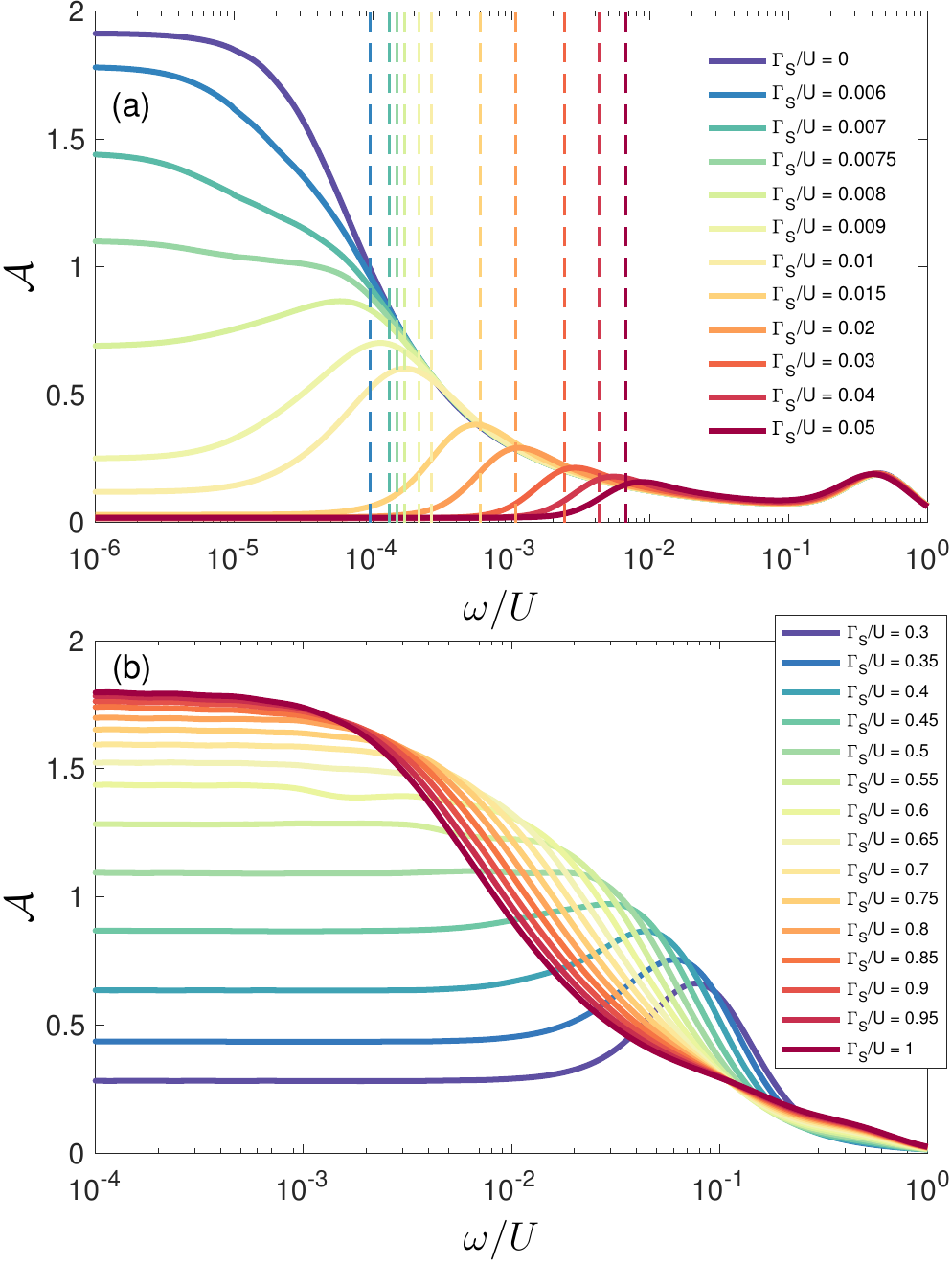}
\caption{\label{Fig:ASU2log}
The total normalized spectral function $\A$ plotted
vs energy on logarithmic scale for selected values of $\Gamma_S$.
Panel (a) presents the suppression of the Kondo resonance with $\Gamma_S$,
which occurs for the critical value of $\Gamma_S = \Gamma_S^{TS}\approx 0.006 U$,
cf. \eq{Eq:GammaSTS}.
The vertical dashed lines in (a) show the excitation energies
between the singlet and triplet states for given $\Gamma_S$,
cf. \eq{Eq:sidepeaks}.
At these excitation energies side Kondo peaks occur.
Panel (b) presents the restoration of the Kondo effect
when $\Gamma_S \gtrsim \Gamma_S^{SD}$.
The parameters are the same as in \fig{Fig:ASU2}.}
\end{figure}

The value of $\Gamma_S$ at which the Kondo resonance
becomes suppressed can be estimated by
comparing the characteristic energy scales, i.e.
the Kondo temperature and the singlet-triplet excitation energy.
One can then find the value of the coupling to superconductor,
$\Gamma_S^{TS}$, at which the suppression of the Kondo resonance develops
\begin{equation} \label{Eq:GammaSTS}
\Gamma_S^{TS} \approx \frac{1}{2}\sqrt{\TKt (U+U_{LR})}.
\end{equation}
For assumed parameters and recalling that $\TKt /U \approx 10^{-4}$,
one gets $\Gamma_S^{TS}/U \approx 0.006$. This estimate is validated
by NRG calculations of the total normalized spectral
function for small values of $\Gamma_S$,
which is plotted as a function of energy on logarithmic scale in \fig{Fig:ASU2log}(a).
One can clearly see the Kondo peak for $\Gamma_S \ll \Gamma_S^{TS}$
and a gradual decrease of its height with increasing $\Gamma_S$,
until the peak becomes completely suppressed 
for $\Gamma_S \gtrsim \Gamma_S^{TS}$.
The vertical dashed lines in \fig{Fig:ASU2log}(a) mark the energy
of the side Kondo peak as estimated from \eq{Eq:sidepeaks}.
The agreement between this analytical formula and full
numerical calculations is quite satisfactory.

For $\Gamma_S \gtrsim \Gamma_S^{TS}$
and such values of $\Gamma_S$ that the ground state is spin singlet,
the system does not exhibit the Kondo effect at all.
The spectral function reveals then just peaks
at energies corresponding to the Andreev bound states,
see \fig{Fig:ASU2}. When, however, the energies of 
Andreev bound states cross the zero energy
for $\Gamma_S \approx \Gamma_S^{SD}$, with
\begin{equation}
\Gamma_S^{SD} = \frac{U(U+2U_{LR})} {8U_{LR}} 
\end{equation}
(for assumed parameters this happens when $\Gamma_S^{SD}=U/2$),
the doublet state $\ket{D_\s}$ becomes the ground state of the system.
Then, one observes the reemergence of the Kondo resonance.
This is explicitly presented in \fig{Fig:ASU2log}(b),
which shows the total normalized spectral function
plotted on logarithmic energy scale for the corresponding values of $\Gamma_S$.
Note that the Kondo temperature
is now clearly larger compared to the case of $\Gamma_S=0$, cf. Figs. \ref{Fig:ASU2log}(a) and (b).
This basically results from the difference in excitation energies to virtual states
allowing for spin-flip processes driving the Kondo effect.
For $\Gamma_S=0$, the energy is given by the charging energy of each dot,
while for $\Gamma_S\gtrsim \Gamma_S^{SD}$, it
is given by the doublet-singlet excitation energy, which is 
smaller than $U$. Consequently,
there is a larger exchange interaction in the latter case,
which explains the observed difference in Kondo temperatures.

It is also interesting to notice that the maximum value of $\A$ at $\omega=0$
is comparable for $\Gamma_S = 0$ and $\Gamma_S=U$,
and approaches $2$, see \fig{Fig:ASU2log}.
In the former case this limit can be easily
understood since each of the two quantum dots contributes
with the Kondo resonance, such that $\A_L=\A_R \to 1$.
In the latter case, on the other hand, one finds, $\A_L=\A_R \to 1/2$
and $\A_{LR} = \A_{RL} \to 1/2$, cf. \fig{Fig:ASU2}, which implies that the off-diagonal
spectral function, that encompasses cross-correlations between the two dots,
contributes $1/(\pi\Gamma)$ to the height of the Kondo peak in the total spectral function.

\begin{figure}[tb]
\centering
\includegraphics[width=0.95\columnwidth]{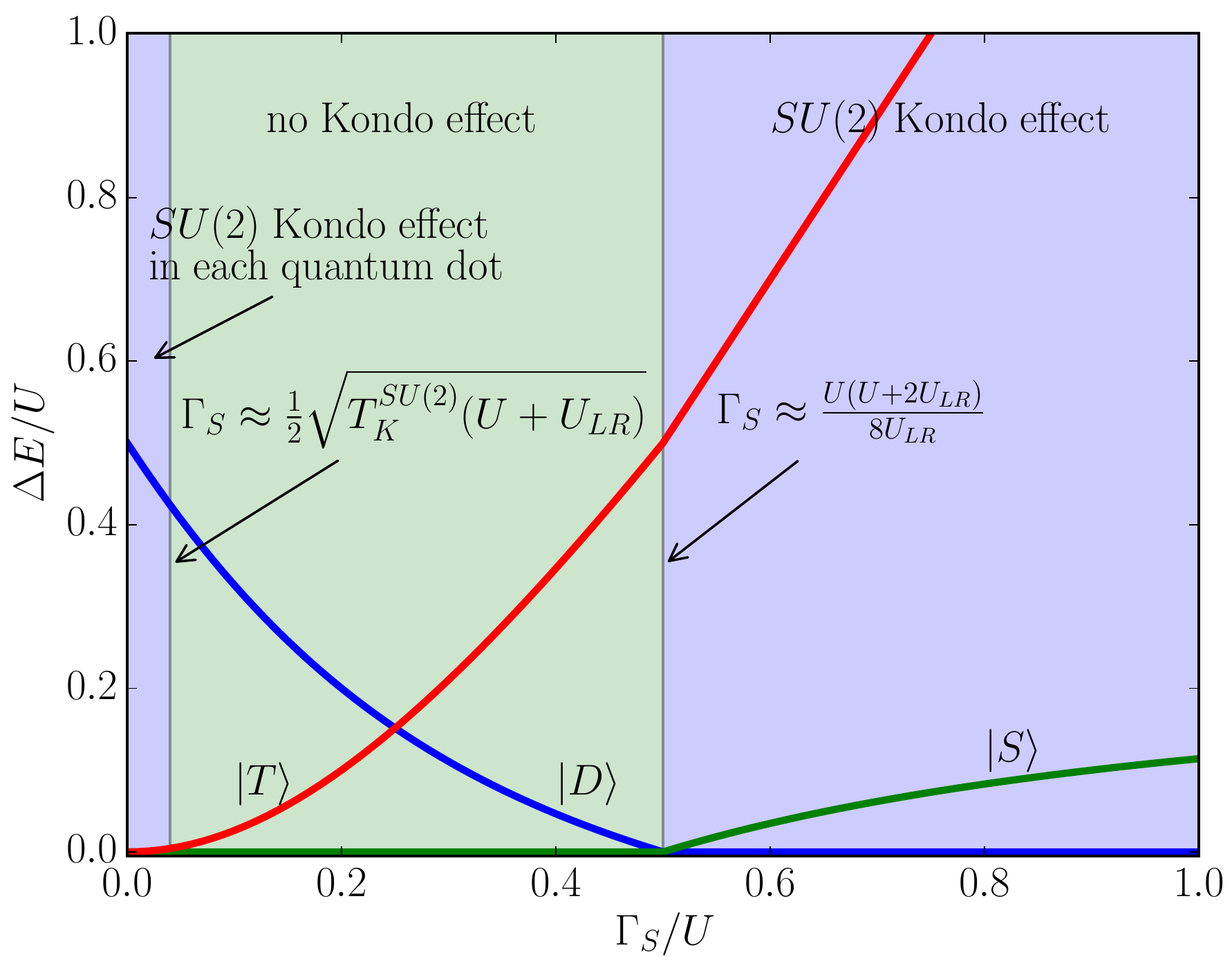}
\caption{\label{Fig:SU2}
The excitation energies $\Delta E$ between the singlet, doublet and triplet states plotted as a function
of the coupling to superconductor for parameters the same as in \fig{Fig:ASU2}.
The excitation energies are measured relative to the ground state
energy, which is set to zero. The evolution of the ground state
from the singlet ($\ket{S}$) to the doublet ($\ket{D}$) state is clearly visible.
The values of $\Gamma_S$ at which the Kondo effect
becomes suppressed or emerges are indicated.
$\TKt$ denotes the $SU(2)$ Kondo temperature for $\Gamma_S =0$.
Note that for $\Gamma_S=0$ the singlet and triplet ($\ket{T}$) 
states are degenerate.}
\end{figure}

The low-energy behavior of the system in the two-electron transport regime
is summarized in \fig{Fig:SU2}, which shows the evolution of the excitation energies $\Delta E$ between the
relevant states, cf. \eq{Eq:en}, when $\Gamma_S$ is varied.
For two indicated values of $\Gamma_S$,
the transport behavior of the system greatly changes.
When $\Gamma_S \lesssim \Gamma_S^{TS}$,
the Kondo singlet is the ground state of the system
and the electrons experience a $\pi/2$ phase shift \cite{Hewson_book}.
At $\Gamma_S \approx \Gamma_S^{TS}$,
there is a crossover, such that for $\Gamma_S^{TS} \lesssim \Gamma_S \lesssim \Gamma_S^{SD}$,
the inter-dot pairing-induced singlet becomes the ground state of the system.
Consequently, there is no Kondo effect (phase shift is equal to $0$).
On the other hand, when $\Gamma_S \approx \Gamma_S^{SD}$,
the system exhibits a phase transition and
for $\Gamma_S \gtrsim \Gamma_S^{SD}$
the doublet state becomes the ground state of the splitter.
This results in the reemergence of the Kondo effect.

Note that the system's behavior as a function
of $\Gamma_S$ is completely different from the case of
a single quantum dot. In single quantum dots
attached to superconducting and normal leads,
in the sub-gap transport regime, the increase of $\Gamma_S$
results in an enhancement of the Kondo temperature
\cite{Domanski2016Mar}.
Since in the case of DQD for $\Gamma_S=0$
the Kondo effect develops on each quantum dot, one could
naively expect that for finite $\Gamma_S$
the behavior will be qualitatively the same as in the single quantum dot case.
The above-presented analysis clearly demonstrates that such
conjecture is completely unjustified.
The proximity-induced inter-dot pairing potential
spoils this picture and, once $\Gamma_S \gtrsim \Gamma_S^{TS}$,
it immediately results in the suppression
of the Kondo resonance on both quantum dots.
Thus, the coupling to superconductor has a strong {\em destructive}
influence on the $SU(2)$ Kondo effect in DQD-based Cooper pair splitters.
Note also that very large value of the coupling $\Gamma_S$, i.e. 
$\Gamma_S \gtrsim \Gamma_S^{SD}$,
can induce the Kondo effect again.

\begin{figure}[t]
\centering
\includegraphics[width=1\columnwidth]{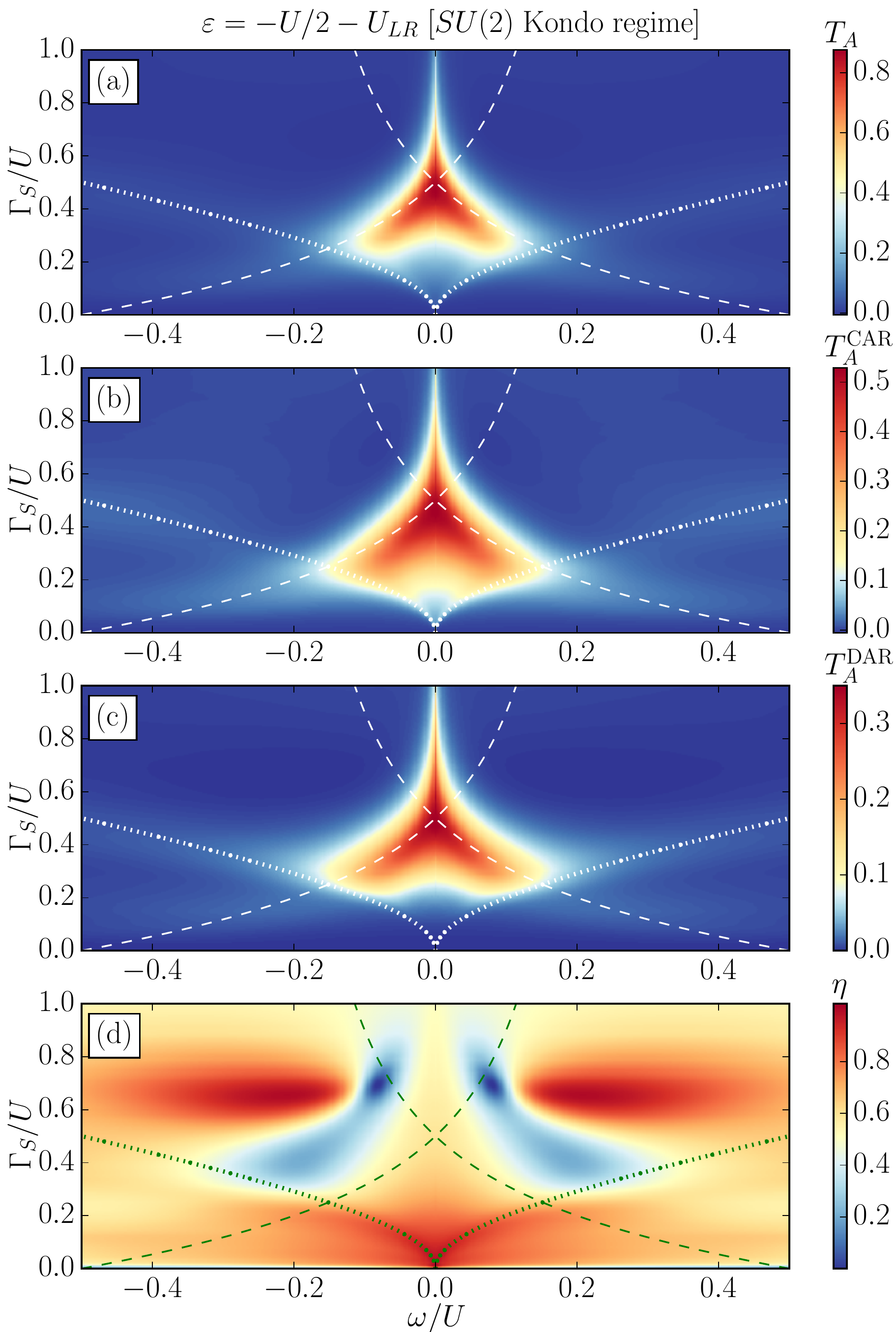}
\caption{\label{Fig:TASU2}
(a) The total Andreev transmission coefficient
and its contributions due to (b) CAR and (c) DAR
processes, as well as (d) the Cooper pair splitting efficiency $\eta$
plotted as function of energy $\omega$
and the strength of coupling to superconductor $\Gamma_S$.
The dashed lines indicate the energies of the Andreev bound states
given by \eq{eq:ABSsu2}, while the dotted lines present
the excitation energies between corresponding singlet and triplet states
given by \eq{Eq:sidepeaks}.
The parameters are the same as in \fig{Fig:ASU2}.
}
\end{figure}

Let us now analyze the behavior of the Andreev transmission,
its contributions due to DAR and CAR processes,
and the splitting efficiency in the $SU(2)$ Kondo regime.
The dependence of these quantities on energy and
strength of coupling to superconductor is presented
in \fig{Fig:TASU2}. First of all, one can see
that the transmission coefficient achieves considerable
values mainly in the low-energy regime, in-between 
the Andreev bound states. Moreover, an enhancement 
of transmission can be also seen along the energies
of Andreev bound states, see \fig{Fig:TASU2}.
Interestingly, we note that for small values of $\Gamma_S$
and low energies, mainly CAR processes dominate transport,
which results in almost perfect splitting efficiency, see 
\fig{Fig:TASU2}(d). We recall that this is the regime 
of suppressed and split Kondo resonance, which now we can clearly
associate with the inter-dot pairing generated by crossed Andreev reflection.
Note that despite suppression of the Kondo effect,
in this transport regime $\TAC$ is still considerable and extends
to energy regions greater than $\TKt$.
When $\Gamma_S \gtrsim \Gamma_S^{SD}$,
at low energies the splitting efficiency is smaller and it indicates
that CAR and DAR processes contribute to Andreev transport 
on an equal footing. On the other hand, for larger energies, $\eta$
first becomes suppressed and then increases again.
However, in this transport regime the total transmission is relatively low,
see \fig{Fig:TASU2}.

\subsection{The SU(4) Kondo regime}

\begin{figure}[t]
\centering
\includegraphics[width=1\columnwidth]{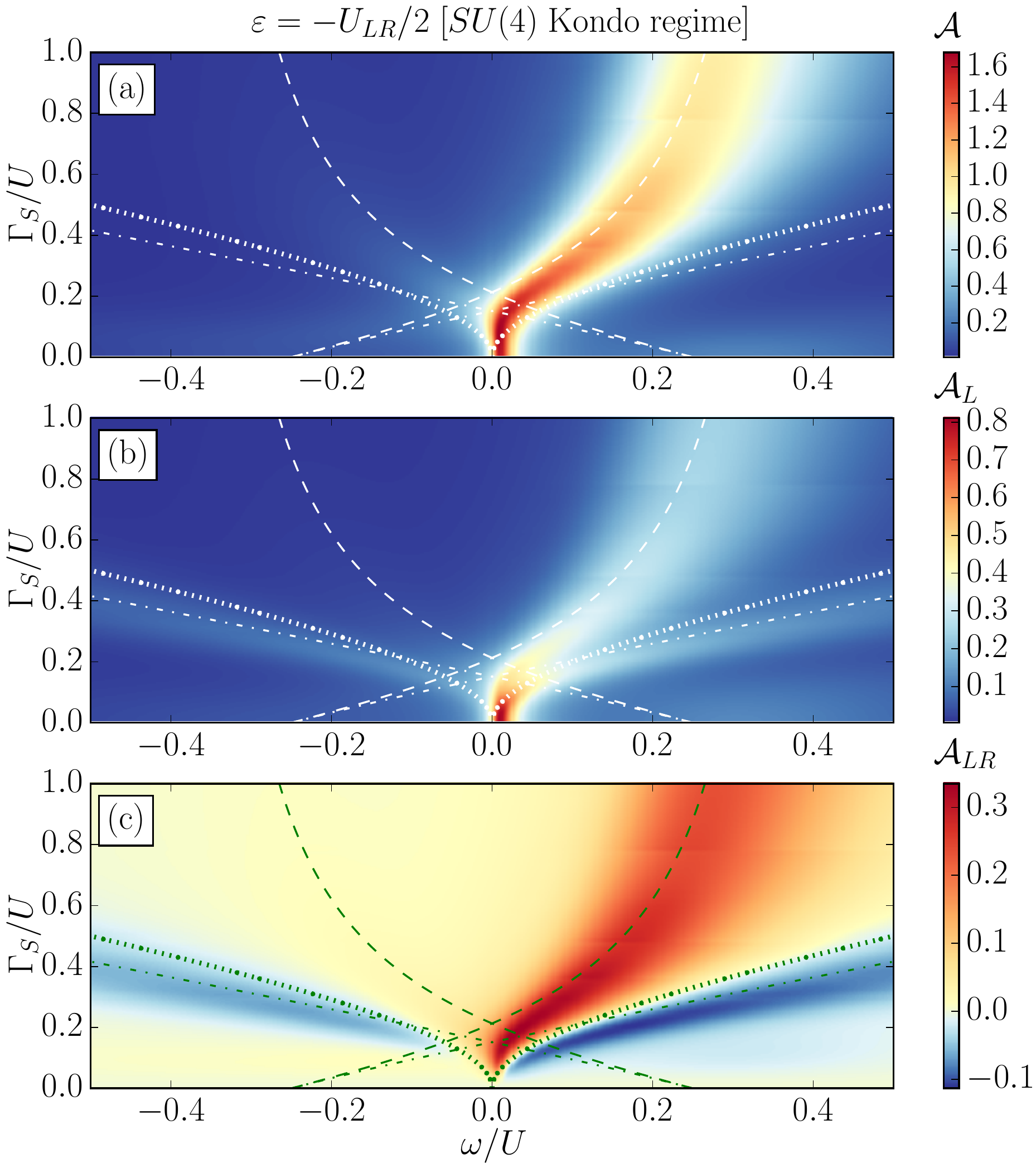}
\caption{\label{Fig:ASU4}
The energy dependence of (a) the total normalized spectral function $\A$
and its contributions: (b) $\A_L$ and (c) $\A_{LR}$
calculated as a function of the coupling to superconductor $\Gamma_S$ and for $\e =-U_{LR}/2$. 
The dashed and dotted-dashed lines indicate the energies of the Andreev bound states,
while the dotted line shows the splitting of the doublet states,
as given by \eq{eq:DeltaEd}.
The other parameters are as in \fig{Fig:AGammaS}.}
\end{figure}

In this section we consider more thoroughly the behavior of the spectral function and
Andreev transmission in the $SU(4)$ Kondo regime, see also \fig{Fig:stab}.
For $\Gamma_S = 0$ and when the DQD is singly occupied,
the system exhibits the $SU(4)$ Kondo effect
resulting from the spin and orbital degeneracies.
For the present analysis we thus assume $\e = -U_{LR}/2$.
The normalized spectral function calculated as a function
of energy and the strength of coupling to superconductor is shown in \fig{Fig:ASU4}.
At first sight, one can deduce that for relatively low values of $\Gamma_S$,
i.e. $\Gamma_S \lesssim U/5$,
the $SU(4)$ Kondo resonance is hardly affected by
the superconducting proximity effect.
Only when the coupling to superconductor 
becomes larger ($\Gamma_S \gtrsim U/5$), does the 
Kondo phenomenon get suppressed---the resonance
in the spectral function becomes then broadened and departs
to larger energies. In fact, for 
$\Gamma_S \approx U/5$, the ground state of the system
changes from the spin doublet to spin singlet state,
and this is the reason for vanishing of the Kondo effect.
For $\Gamma_S \gtrsim U/5$, $\A$
exhibits only resonances at larger energies corresponding to the Andreev
bound state energies, see the dashed and dotted-dashed lines in \fig{Fig:ASU4},
which mark the energies of Andreev bound states.
The ABS's energies were determined from the excitation energies
between appropriate singlet and doublet states
obtained from numerical solution of the eigenvalue problem.
The resonances associated with excitations due to Andreev bound states
are also clearly visible in the spectral function of individual quantum dots $\A_L$
as well as in $\A_{LR}$, shown in Figs. \ref{Fig:ASU4}(b) and (c), respectively.

\begin{figure}[t]
\centering
\includegraphics[width=1\columnwidth]{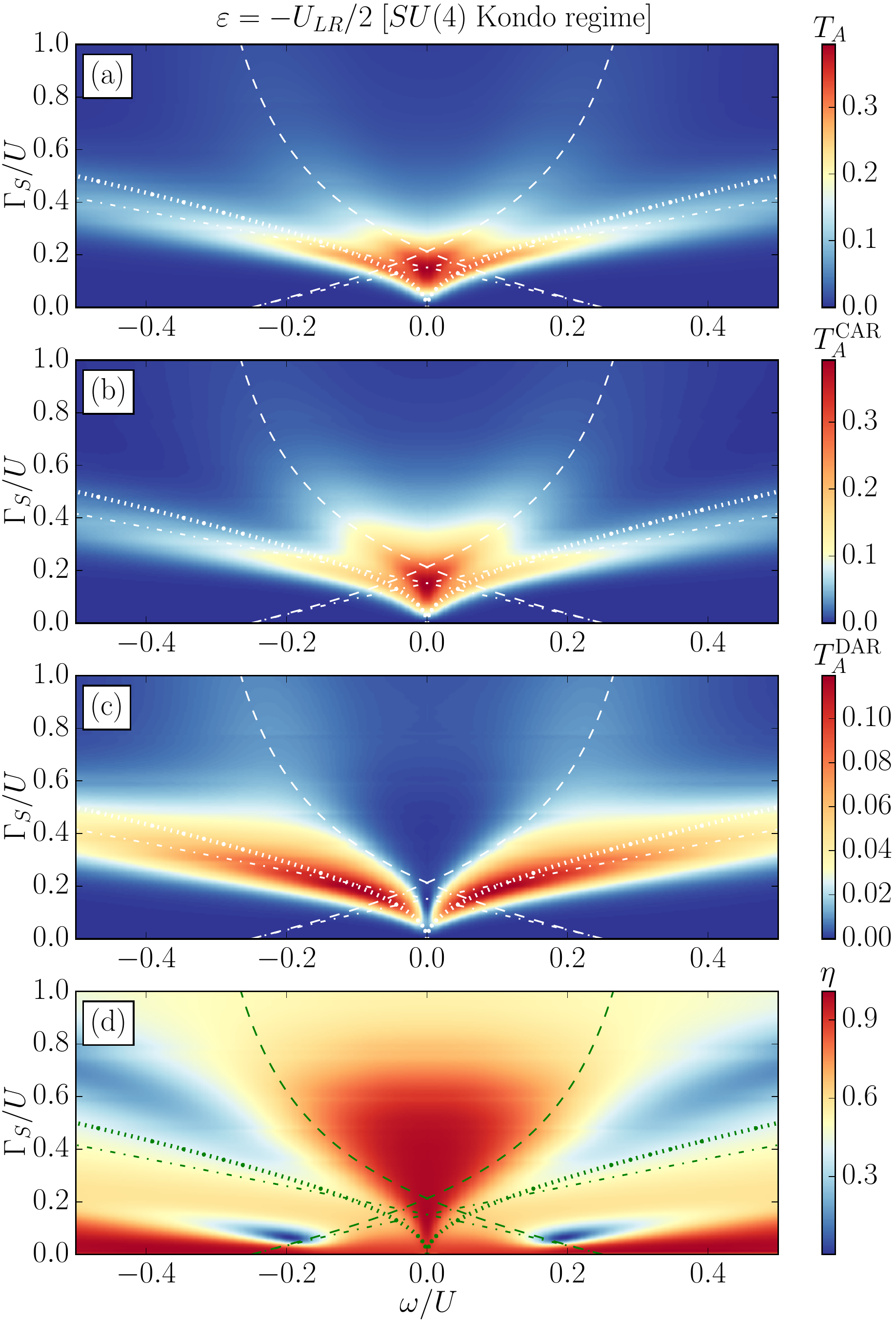}
\caption{\label{Fig:TASU4}
(a) The total Andreev transmission coefficient
and its contributions due to (b) CAR and (c) DAR
processes, as well as (d) the Cooper pair splitting efficiency $\eta$
plotted as function of energy $\omega$
and the strength of coupling to superconductor $\Gamma_S$
for parameters the same as in \fig{Fig:ASU4}.
The dashed and dotted-dashed lines indicate the energies of the Andreev bound states,
and the dotted line shows the splitting of the doublet states, as described by \eq{eq:DeltaEd}.}
\end{figure}

At energies corresponding to Andreev bound states,
the Andreev transmission coefficient also becomes enhanced.
This can be seen in \fig{Fig:TASU4}, which presents
the energy $\omega$ and $\Gamma_S$ dependence
of $\TA$ and its contributions due to CAR and DAR processes,
together with the splitting efficiency $\eta$.
We again notice that generally $\TAC > \TAD$
[cf. Figs. \ref{Fig:TASU4}(b) and (c)],
which leads to large splitting efficiency, especially 
visible for low energies [\fig{Fig:TASU4}(d)].
In fact, for $\Gamma_S \approx U/5$,
i.e. when the doublet-singlet transition occurs, the 
total transmission coefficient has a local maximum,
which results mainly from CAR processes, see \fig{Fig:TASU4}.
Consequently, the splitting efficiency becomes then very close to unity.
On the other hand, there are also transport regimes where $\eta$
is very much suppressed, which indicates that DAR processes are dominant,
see the transport region for $|\omega|\approx U/5$ and low values of $\Gamma_S$
($\Gamma_S \approx U/10$) in \fig{Fig:TASU4}(d).
The transmission coefficient in these transport regimes
is however relatively low.

\begin{figure}[t]
\centering
\includegraphics[width=1\columnwidth]{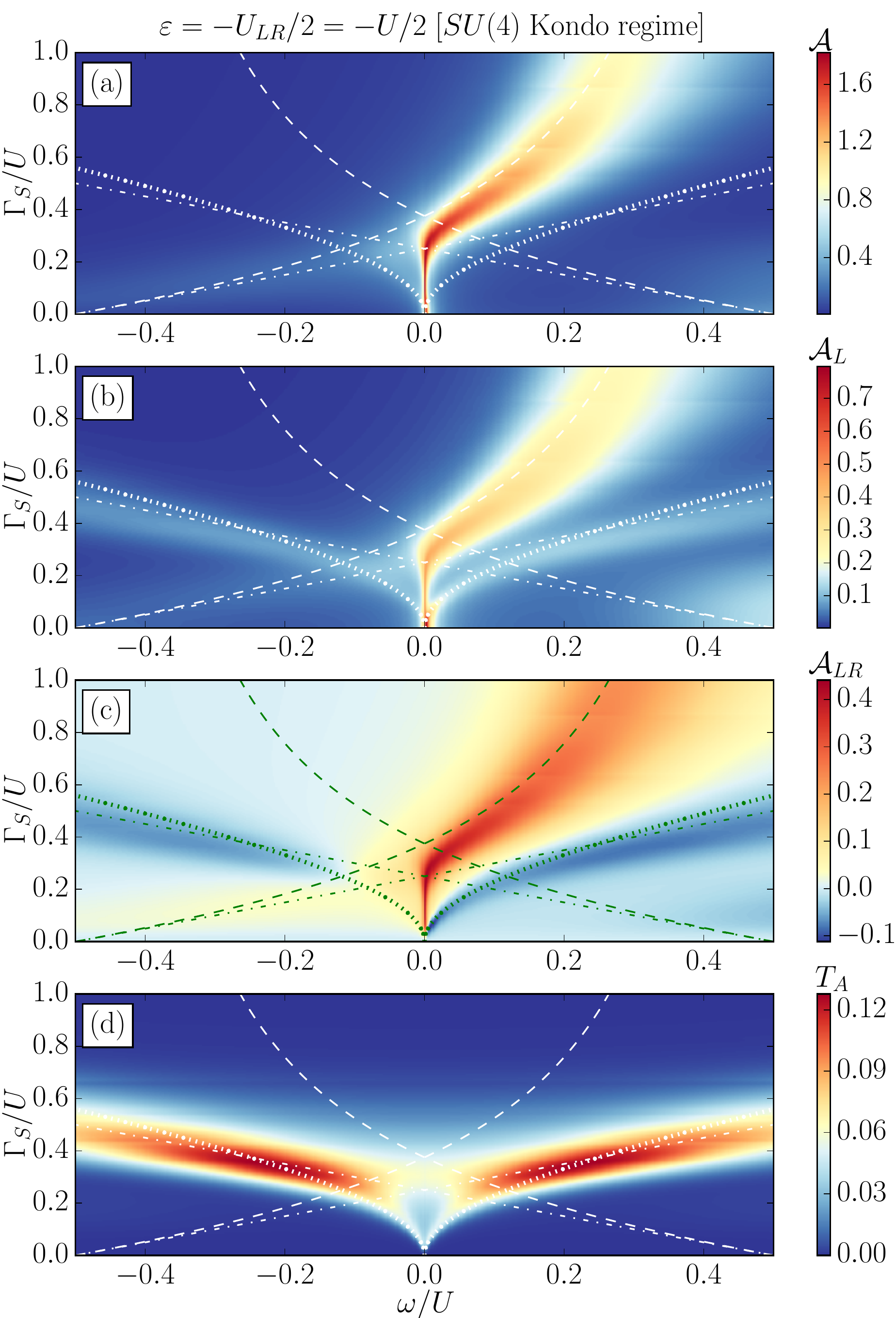}
\caption{\label{Fig:ASU4sym}
The energy dependence of (a) the total normalized spectral function $\A$,
its contributions (b) $\A_L$ and (c) $\A_{LR}$,
and (d) the total Andreev transmission coefficient $\TA$
plotted as a function of energy $\omega$ and $\Gamma_S$
in the $SU(4)$ Kondo regime for the symmetric case with $U_{LR}=U$.
The dotted-dashed (dashed) lines indicate the energies of the Andreev bound states
$E_{ABS}^1$ ($E_{ABS}^2$), cf. \eq{eq:ABSsu4}, while the dotted
lines show the doublet splitting energy, cf. \eq{eq:DeltaEd}.
The other parameters are the same as in \fig{Fig:AGammaS}.}
\end{figure}

To shed more light on the influence of superconducting pairing
correlations on the $SU(4)$ Kondo regime,
let us now assume a fully symmetric situation,
namely $U_{LR} = U$.
For this case, the dependence of the relevant spectral functions
and the total Andreev transmission coefficient on $\omega$ and $\Gamma_S$
is shown in \fig{Fig:ASU4sym}.
In the symmetric case, one can find the 
eigenenergies and eigenstates in the spin singlet subspace explicitly.
These are presented in Table \ref{tab3} in the Appendix,
while the eigenspectrum in the doublet subspace can be found in Table \ref{tab1}.
Note that in the doublet subspace we can find the eigenspectrum
for arbitrary parameters, therefore, if only the doublet states are considered
we will present the analytical formulas for general case of $U_{LR}\neq U$.
From the inspection of the spectrum of $H_{DQD}^{\rm eff}$ one can see
that for $\Gamma_S \to 0$ the ground state is indeed four-fold degenerate
and given by the doublet states
\begin{equation}
\ket{D_\s^1} = \frac{1}{\sqrt{2}} (\ket{\s0}-\ket{0\s})
\end{equation}
with energy $E_D^1=-U_{LR}/2$ and
\begin{equation}
\ket{D_\s^2} \!=\! \frac{1}{\sqrt{16\G_S^2\!+\!\alpha^2}}\big[
\alpha(\ket{\s0}\!+\!\ket{0\s}) + 4\G_S(\ket{\s d}\!+\!\ket{d\s})
\big],
\end{equation}
with $\alpha = U+U_{LR}+\sqrt{(U+U_{LR})^2+16\G_S^2}$, and the energy,
$E_D^2 = U/2 - \sqrt{(U+U_{LR})^2 + 16\Gamma_S^2}/2$.
With increasing $\Gamma_S$, the two doublet states become split
and the ground state is given by the state $\ket{D_\s^2}$.
The doublet splitting energy is given by
\begin{equation}
\label{eq:DeltaEd}
\omega = \pm \frac{1}{2} \bigg[U+U_{LR} - \sqrt{(U+U_{LR})^2+16\G_S^2}\bigg].
\end{equation}
This energy difference is marked with dotted lines
in Figs. \ref{Fig:ASU4}, \ref{Fig:TASU4}, \ref{Fig:ASU4sym}.
It coincides with the resonances in the spectral
function $\A_L$ obtained from NRG calculations.
These resonances are however not visible in the total spectral function,
since the peak in $\A_L$ is counterbalanced by an associated minimum in 
$\A_{LR}$, see e.g. Figs. \ref{Fig:ASU4}(b) and (c).
Pronounced maxima can be also observed in the 
Andreev transmission coefficient shown in Figs. \ref{Fig:TASU4} and \ref{Fig:ASU4sym}(d).
Note that while around the Fermi energy both the spectral function
and Andreev transmission show features at the doublet-doublet excitation
energy [\eq{eq:DeltaEd}], for larger $\omega$, the resonances
occur at energies corresponding rather to the Andreev bound states.

\begin{figure}[t]
\centering
\includegraphics[width=0.95\columnwidth]{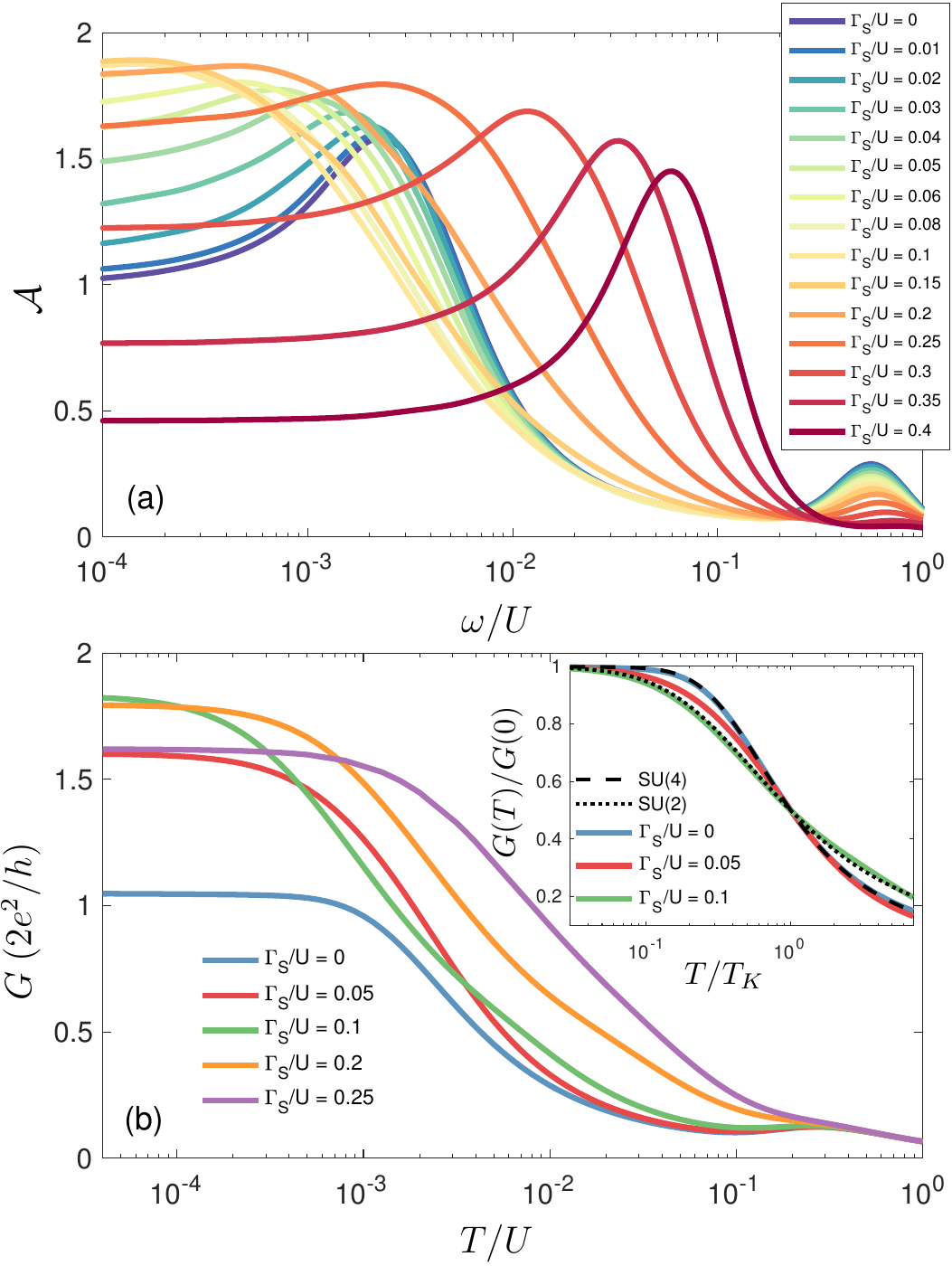}
\caption{\label{Fig:ASU4symlog}
(a) The total spectral function plotted vs energy
on logarithmic scale and 
(b) the temperature dependence 
of the linear-response normal conductance
for different values of $\Gamma_S$,
as indicated in the legends.
The inset in (b) presents the crossover of the universal scaling 
of the conductance vs $T/T_K$ from the $SU(4)$
to the $SU(2)$ Kondo regime.
The other parameters are the same as in \fig{Fig:ASU4sym}.
For $\Gamma_S=0$ and assumed parameters
$\TKf /U\approx 0.004$.}
\end{figure}

The influence of the superconducting pairing correlations on the $SU(4)$ Kondo state
can be better resolved in the spectral function plotted versus energy
on logarithmic scale. This is presented in \fig{Fig:ASU4symlog}.
Now, one can clearly see that the maximum in the spectral function
strongly depends on $\Gamma_S$. For very small pairing correlations,
$\A$ exhibits a resonance at finite $\omega$, see \fig{Fig:ASU4symlog}.
Now for assumed parameters and $\Gamma_S=0$
one finds $\TKf /U\approx 0.004$.
However, with increasing $\Gamma_S$, this resonance becomes suppressed
and moves towards the Fermi energy. This is a clear indication
of a crossover from the $SU(4)$ to the $SU(2)$ Kondo effect.
Finite pairing correlations break the four-fold degeneracy of the ground state
and reduce it to two-fold degeneracy due to only the spin degrees of freedom.
Because of that, the $SU(4)$ Kondo effect becomes suppressed.
One can estimate the strength of coupling $\Gamma_S$
when this crossover takes place ($\Gamma_S^{DD}$)
by comparing the doublet splitting energy with the corresponding Kondo temperature.
This leads to
\begin{equation} \label{Eq:GammaSTK4}
\Gamma_S^{DD} \approx \frac{1}{2}\sqrt{\TKf (U+U_{LR})}.
\end{equation}
For parameters assumed in \fig{Fig:ASU4symlog}
one then finds $\Gamma_S^{DD}\approx 0.045 U$.
This estimate agrees reasonably well with 
the numerical data shown in \fig{Fig:ASU4symlog}(a).

Moreover, we corroborate the $SU(4)$-$SU(2)$ crossover by calculating
the temperature dependence of the normal conductance,
which for potential drop between the left and right leads
in the linear response regime can be expressed as \cite{Trocha2014_PhysRevB.89.245418,Bocian2017Apr},
\beq \label{Eq:G}
  G = \frac{2e^2}{h} \int \!\! d\omega  \left(-\frac{\partial f(\omega)}{\partial \omega} \right) \A,
\eeq
where $f(\omega)$ is the Fermi-Dirac distribution function.
We note that, since the total conductance may contain contributions
from Andreev reflection processes, the normal conductance $G$
should be considered as a theoretical
tool to gain information about the type of scaling
and, thus, the type of the Kondo effect in the system.
The temperature dependence of $G$ is shown in \fig{Fig:ASU4symlog}(b).
For $\Gamma_S = 0$, $G(T)$ exhibits the $SU(4)$
universal scaling, see the inset in \fig{Fig:ASU4symlog}(b).
However, with increasing $\Gamma_S$,
e.g. for $\Gamma_S = 0.05 U \gtrsim \Gamma_S^{DD}$,
the scaling does not collapse onto the $SU(4)$ universal function
any more. Instead, for $\Gamma_S > \Gamma_S^{DD}$,
one finds that the $SU(2)$ Kondo effect becomes 
responsible for the conductance enhancement.
The conductance reveals the $SU(2)$ universal scaling
for $\Gamma_S$ up to $\Gamma_S\approx 3U/10$ (not shown),
since for larger $\Gamma_S$ the doublet 
is not the ground state of the system any more
and the Kondo effect is not present in the system.

\begin{figure}[tb]
\centering
\includegraphics[width=0.95\columnwidth]{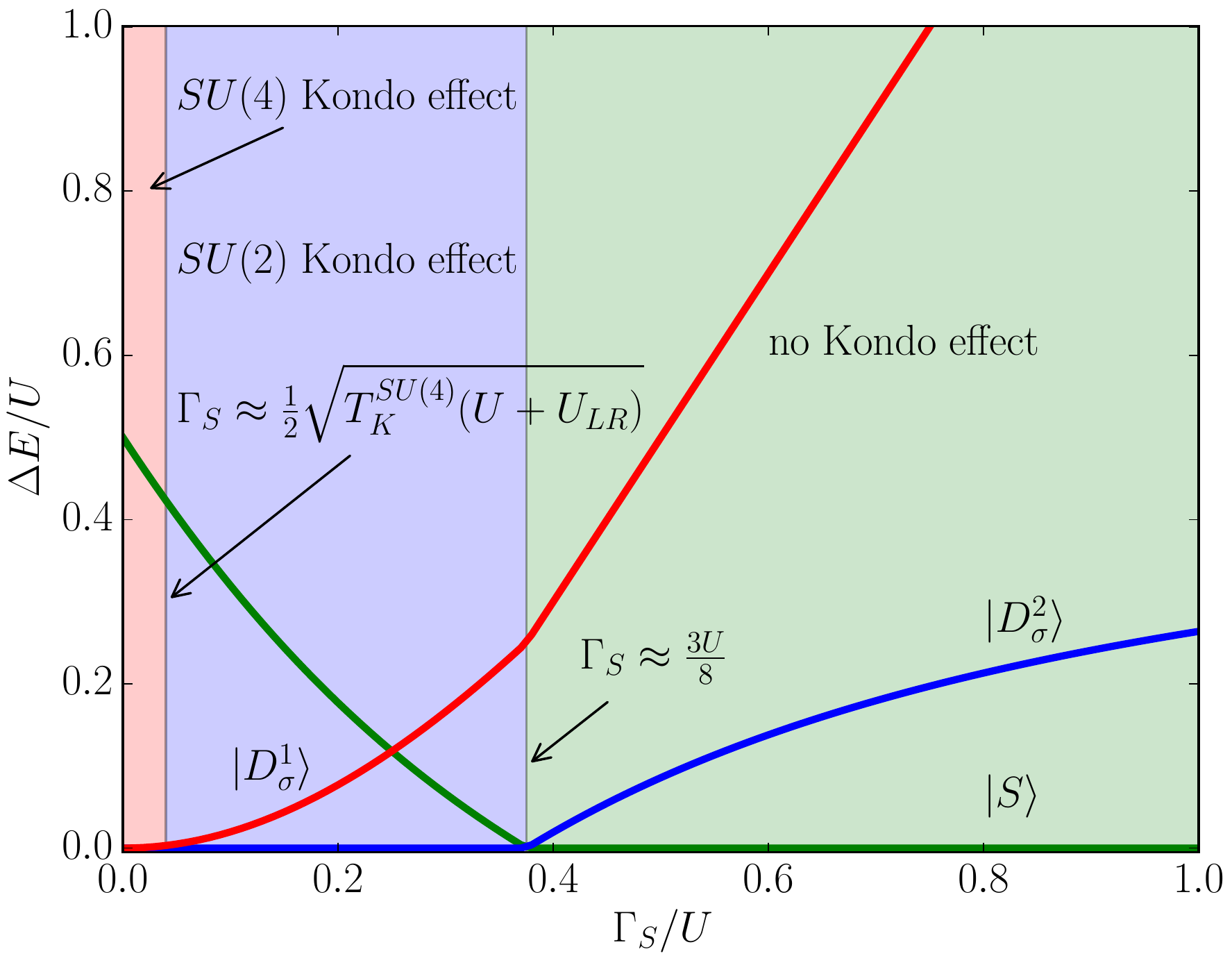}
\caption{\label{Fig:SU4}
The excitation energies $\Delta E$ between the corresponding
singlet and two doublet states plotted as a function
of the coupling to superconductor for parameters the same as in \fig{Fig:ASU4sym}.
The excitation energies are measured relative to the ground state
energy, which is set to zero. 
The values of $\Gamma_S$ at which the symmetry of the Kondo state
or the ground state of the system changes are indicated.}
\end{figure}

Note also that the maximum value of the low-temperature conductance,
which corresponds directly to $\A$ at $\omega=0$,
depends in a nonmonotonic fashion on $\Gamma_S$.
For $\Gamma_S = 0$, $G=2e^2/h$, while 
for $\Gamma_S^{DD}\lesssim \Gamma_S$,
$G$ is clearly larger than $2e^2/h$ and approaches almost 
$4e^2/h$, see \fig{Fig:ASU4symlog}(b).
This can be understood by realizing that finite coupling to superconductor
leads to an enhancement of the average occupation of each dot,
such that the occupation of the double dot becomes larger than one.
Moreover, finite coupling to superconductor results
in a large enhancement of $\A_{LR}$, such that the total
conductance reaches $G\approx 4e^2/h$.

When the coupling to superconductor is enhanced further,
a doublet-singlet transition occurs for $\Gamma_S = 3U/8$.
For $\Gamma_S > 3U/8$, the ground state of the system
is given by the following singlet state (cf. state $\ket{S_2}$ in Table \ref{tab3})
\begin{equation}
\ket{S} \!=\! \frac{1}{\sqrt{2}}\Big[\ket{00} + \frac{1}{2}(\ket{d0} \!+\! \ket{0d})
+ \frac{1}{2}(\ket{\up\down} \!-\! \ket{\down\up})\Big] ,
\end{equation}
with the energy $E_S = -2\Gamma_S$.
The excitations between the singlet and the two doublet states
allow us to estimate the analytical formulas for the energies
of the relevant Andreev bound states, which are given by
\begin{eqnarray} \label{eq:ABSsu4}
E_{ABS}^1 &=& \pm\frac{U}{2}\mp 2\G_S,\nonumber \\
E_{ABS}^2 &=& \pm\frac{U}{2}\pm 2\G_S\mp \sqrt{U^2+4\G_S^2}.
\end{eqnarray}
The energies of those Andreev bound states
are presented in \fig{Fig:ASU4sym} with dotted-dashed and dashed lines, respectively.
In fact, for $\Gamma_S > 3U/8$, the resonances
present in the spectral function for positive energies are exactly due
to the Andreev bound states, see \fig{Fig:ASU4sym}.
At the ABS energy $E_{ABS}^1$ an enhancement of 
the Andreev transmission is also clearly present, 
see \fig{Fig:ASU4sym}(d).

Summing up, in the $SU(4)$ Kondo regime,
i.e. for $\e=-U/2$ with $U_{LR} = U$,
the $SU(4)$ Kondo effect is present for $\Gamma_S \lesssim \Gamma_S^{DD}$.
At $\Gamma_S\approx \Gamma_S^{DD}$, there is
an $SU(4)$-$SU(2)$ crossover, and for $\Gamma_S^{DD}\lesssim \Gamma_S \lesssim 3U/8$
the system exhibits the $SU(2)$ Kondo resonance.
When $\Gamma_S \approx 3U/8$, there is a phase transition
and the ground state changes from the spin doublet to the spin singlet state,
such that for larger values of $\Gamma_S$
the system does not exhibit the Kondo effect any more.
These findings are schematically summarized in \fig{Fig:SU4},
which shows the evolution of the ground state when the strength of coupling
to superconductor increases.

\section{Conclusions}

We have analyzed the transport properties of double quantum dot based
Cooper pair splitters strongly coupled to external electrodes, focusing on the Kondo regime.
The two dots were attached to a common $s$-wave superconductor and
each dot was coupled to a separate metallic electrode.
The considerations were performed in the subgap transport regime,
where transport was driven by direct and crossed Andreev reflection processes.
By using the density-matrix numerical renormalization group method,
we determined the behavior of the local density of states of DQD
and the Andreev transmission coefficient, together with Cooper-pair splitting 
efficiency. First, we have analyzed the dependence of the 
transport properties on the position of the DQD energy levels
and then we have focused on the $SU(2)$ and $SU(4)$ Kondo regimes.

We have shown that the superconducting pairing correlations
can greatly influence the Kondo effect in the system.
In the $SU(2)$ Kondo regime, we predict
a very quick suppression of the Kondo resonance
with increasing the strength of coupling to superconductor.
This effect is in stark contrast to single quantum dot case,
where increase of pairing correlations resulted in
an enhancement of the Kondo temperature \cite{Zitko2015Jan,Domanski2016Mar}.
The disappearance of the $SU(2)$ Kondo peak
is directly associated with the formation of a spin singlet state
between the two quantum dots triggered by 
proximity-induced inter-dot pairing potential.
With increasing the strength of coupling to superconductor further,
we demonstrate that the system undergoes a transition to the doublet state.
In this transport regime, the Kondo effect reemerges
and the total spectral function shows a pronounced Kondo peak.
The occurrence of this resonance
is associated with  contributions coming from both individual quantum dots $\A_i$,
as well as from cross-correlations described by off-diagonal part
of the spectral function $\A_{ij}$.

In the $SU(4)$ Kondo regime, on the other hand,
the impact of superconducting pairing correlations on the Kondo state
is less abrupt and now the Kondo effect persists for 
larger couplings to superconductor as compared to the $SU(2)$ case.
More specifically, we predict that,
in the fully symmetric situation, the $SU(4)$ Kondo effect becomes first reduced
to the $SU(2)$ Kondo effect, which becomes then
fully suppressed once $\Gamma_S > 3U/8$.
For this value of coupling to superconductor, 
the ground state changes from the spin doublet to proximity-induced
singlet state and, consequently, there is no Kondo effect.
The spectral function exhibits then only
resonances at energies corresponding to 
energies of Andreev bound states.
Interestingly, in the $SU(4)$ Kondo regime, when $U_{LR}<U$,
we find that the Andreev current is mainly due to CAR processes,
which yields almost perfect Cooper pair splitting efficiency.

Finally, we would like to note that most of our findings,
and especially the suppression or reemergence of the Kondo state
as the coupling to superconductor is varied,
could be tested with the present-day experimental technology.
We hope that our research will stimulate further efforts in this direction.


\acknowledgements
We acknowledge discussions with Kacper Bocian.
This work was supported by the National Science Centre
in Poland through the Project No. DEC-2013/10/E/ST3/00213.
Computing time at Pozna\'n Supercomputing and Networking Center is acknowledged.


\appendix

\section{\label{app:EigSys} Spectrum of the effective double dot Hamiltonian}

Here we present the eigenvalues and eigenvectors of isolated
double quantum dot with proximity-induced pairing potentials,
as modeled by the effective Hamiltonian (\ref{Eq:HDQDeff}).
Because the Hamiltonian possesses the full spin $SU(2)$ symmetry,
we can write $H_{DQD}^{\rm eff}$ in blocks
labeled by the spin quantum number.
Moreover, it is enough to use $10$ spin multiplets instead of $16$ local states.
Let us first start from the trivial triplet subspace. The triplet state $\ket{T_\d}$
has the components:
$\ket{T_{+}}=\ket{\!\up\up}$,
$\ket{T_{-}}=\ket{\!\down\down}$,
$\ket{T_{0}}=(\ket{\!\up\down} + \ket{\!\down\up})/\sqrt{2}$,
and the energy $E_T = 2\e + U_{LR}$.

The Hamiltonian block in the spin doublet subspace
is explicitly given by
\begin{equation}
H_{DQD}^{\rm eff,S=\tfrac{1}{2}} \!=\!\!
\left(\!
\begin{array}{cccc}
  \e & 0 & -\G_S & -\G_S\\
  0 & \e & -\G_S & -\G_S\\  
  -\G_S & -\G_S & \e_3& 0\\
  -\G_S & -\G_S & 0 & \e_3
\end{array}
\!\right),
\end{equation}
with $\e_3 = 3\e+2U_{LR}+U$.
This matrix is written in the following states:
$\ket{\s0}$, $\ket{0\s}$, $\ket{\s d}$ and $\ket{d\s}$, 
respectively, and its eigenvalues together with unnormalized eigenvectors
are listed in Table \ref{tab1}.
\begin{table}[t!]
\caption{\label{tab1}
The eigenvalues and unnormalized eigenvectors
of the effective DQD Hamiltonian in the doublet subspace.
Here, $\Delta_D = [(2\e+2U_{LR}+U)^2+16\G_S^2]^{\tfrac{1}{2}}$ and
$\alpha = (2\e + 2U_{LR}+U+\Delta_D)/(4\G_S)$.
} \centering
\begin{tabular}{lll}
State\;\;& Eigenenergy \;\;\;\;& Eigenvector\\
\hline
$\ket{D_\s^1}$ & $\e$ & $\ket{\s0}-\ket{0\s}$\\
$\ket{D_\s^2}$ & $2\e+U_{LR}+\tfrac{U-\Delta_D}{2}$ & $\alpha (\ket{\s0}+\ket{0\s})+\ket{\s d}+\ket{d\s}$\\
$\ket{D_\s^3}$ & $2\e+U_{LR}+\tfrac{U+\Delta_D}{2}$ & $\ket{\s0}+\ket{0\s}-\alpha (\ket{\s d}+\ket{d\s})$\\
$\ket{D_\s^4}$ & $3\e+2U_{LR}+U$ & $\ket{\s d}-\ket{d\s}$
\end{tabular}
\end{table}

Now, let us consider the singlet subspace which is spanned by the following
five states: $\ket{00}$, $\ket{d0}$, $\ket{0d}$, $\ket{S_0} = (\ket{\up\down}-\ket{\down\up})/\sqrt{2}$
and $\ket{dd}$. The effective DQD Hamiltonian in this subspace is
given by
\begin{equation}
H_{DQD}^{\rm eff,S=0} \!=\!\!
\left(\!\!\!
\begin{array}{ccccc}
  0 & -\G_S & -\G_S & \sqrt{2}\Gamma_S & 0\\
  -\G_S & \! 2\e+U& 0& 0&-\G_S \\  
  -\G_S & 0 & \!\! 2\e+U& 0&-\G_S \\  
  \sqrt{2} \G_S & 0 & 0 & \!\! 2\e+U_{LR}& -\sqrt{2}\G_S \\  
  0 & -\G_S & -\G_S & -\sqrt{2}\G_S & \e_4
\end{array}
\!\!\!\right),
\end{equation}
where $\e_4 = 4\e+2U+4U_{LR}$ is the energy of the fully occupied double dot.
The first eigenstate is: $\ket{S_1} = (\ket{d0} - \ket{0d})/\sqrt{2}$
and its eigenenergy reads $2\e+U$. The next eigenenergies
are given by polynomials of various
Hamiltonian parameters and do not have simple analytical structure,
therefore, we will not present them here.
Instead, let us consider some limiting situations.
The first one is relevant to the $SU(2)$ Kondo regime,
$\e = -U/2-U_{LR}$, and the second one is
associated with the $SU(4)$ Kondo regime,
when $\e = -U_{LR}/2$ and $U=U_{LR}$.
The eigenspectrum in the former case
is presented in Table \ref{tab2},
while the states and energies
in the latter case are listed in Table \ref{tab3}.

\begin{table}[t!]
\caption{\label{tab2}
The eigenvalues and unnormalized eigenvectors
of the effective DQD Hamiltonian in the singlet subspace
for the particle-hole symmetry point, $\e=-U/2-U_{LR}$.
Here, $\Delta_S = (U_{LR}^2+4\G_S^2)^{\tfrac{1}{2}}$,
$\alpha = (U_{LR}+\Delta_S)/(2\G_S)$ and
$\tilde \Delta_S = [(U+U_{LR})^2+16\G_S^2]^{\tfrac{1}{2}}$.
} \centering
\begin{tabular}{lll}
State\;\;\;\;& Eigenenergy \;\;\;\;& Eigenvector\\
\hline
$\ket{S_1}$ & $-2U_{LR}$ & $\ket{d 0}-\ket{0d}$\\
$\ket{S_2}$ & $-U_{LR}-\Delta_{S}$ & $\ket{dd}+\ket{00} + \alpha (\ket{d 0}+\ket{0d})$\\
$\ket{S_3}$ & $-U_{LR}+\Delta_{S}$ & $\alpha(\ket{dd}+\ket{00}) - \ket{d 0}-\ket{0d}$\\
$\ket{S_4}$ & $-\frac{U+U_{LR}+\tilde\Delta_S}{2}$ & $\ket{dd}-\ket{00}+ \frac{U+U_{LR}+\tilde\Delta_S}{2\sqrt{2}\G_S} \ket{S_0}$\\
$\ket{S_5}$ & $-\frac{U+U_{LR}-\tilde\Delta_S}{2}$ & $\ket{dd}-\ket{00}+ \frac{U+U_{LR}-\tilde\Delta_S}{2\sqrt{2}\G_S} \ket{S_0}$
\end{tabular}
\end{table}

\begin{table}[t!]
\caption{\label{tab3}
The eigenvalues and unnormalized eigenvectors
of the effective DQD Hamiltonian in the singlet subspace
in the $SU(4)$ Kondo regime, that is for $\e=-U/2$
and $U_{LR} = U$.
Here, $\Delta_S = (U^2+\G_S^2)^{\tfrac{1}{2}}$ and
$\alpha = (U+\Delta_S)/(2\G_S)$.
} \centering
\begin{tabular}{lll}
State\;\;\;\;& Eigenenergy \;\;\;\;& Eigenvector\\
\hline
$\ket{S_1}$ & $0$ & $\ket{d 0}-\ket{0d}$\\
$\ket{S_2}$ & $-2\G_S$ & $\ket{S_0}-\sqrt{2}\ket{00} -\frac{1}{\sqrt{2}}(\ket{d 0}+\ket{0d})$\\
$\ket{S_3}$ & $2\G_S$ & $\ket{S_0}+\sqrt{2}\ket{00} -\frac{1}{\sqrt{2}}(\ket{d 0}+\ket{0d})$\\
$\ket{S_4}$ & $2U-2\Delta_S$ & $\alpha(\ket{d0}+\ket{0d}+\sqrt{2}\ket{S_0}) + \ket{dd}$\\
$\ket{S_5}$ & $2U+2\Delta_S$ & $\alpha \ket{dd} - (\ket{d0}+\ket{0d}+\sqrt{2}\ket{S_0})$
\end{tabular}
\end{table}


%

\end{document}